\documentclass{LMCS}

\def\doi{8(3:28)2012}
\lmcsheading%
{\doi}
{1--35}
{}
{}
{Nov.~24, 2009}
{Sep.~29, 2012}
{}
 
\usepackage{enumerate}
\usepackage{hyperref}
\usepackage[english]{babel}
\usepackage[latin1]{inputenc} 
\usepackage[T1]{fontenc}
\usepackage{graphicx}
\usepackage{amssymb}
\usepackage{amsfonts}
\usepackage{amsmath}
\usepackage{verbatim}
\usepackage{url}
\usepackage[usenames]{color}
\usepackage{stmaryrd}
\usepackage{latexsym}
\usepackage{gastex}

\theoremstyle{plain}

\newtheorem{remark}[thm]{Remark}

\newif\ifshort
\shortfalse

\ifshort%
\newcommand\ForAuthors[1]{}%
\else%
 \newcommand\ForAuthors[1]%          %  temporary remark for the
 {\par\smallskip                     %  authors:
  \begin{center}%                    %
   \fbox%                            %    --------
   {\parbox{0.9\linewidth}%          %    |  #1  |
    {\raggedright\sc--- #1}%         %    --------
   }%                                %
  \end{center}%                      %
  \par\smallskip                     %
 }
\fi

\newcommand\mopen{\{\mkern-\thinmuskip|}
\newcommand\mclose{|\mkern-\thinmuskip\}}

\newcommand\Sober{\mathcal S}
\newcommand\Hoare{\mathcal H}

\newcommand\nat{\mathbb{N}}
\newcommand\Z{\mathbb{Z}}
\newcommand\pow{\mathop{\mathbb P}}
\newcommand\Pfin{\pow_{\text{fin}}}
\newcommand\dom{\mathop{\mathrm{dom}}}

\newcommand\limp{\Longrightarrow}

\newcommand\upc{\mathop{\uparrow}\nolimits}
\newcommand\dc{\mathop{\downarrow}\nolimits}
\newcommand\uuarrow{\rlap{$\uparrow$}\raise.5ex\hbox{$\uparrow$}}%%
\newcommand\ddarrow{\rlap{$\downarrow$}\raise.5ex\hbox{$\downarrow$}}%%

\newcommand\accel[1]{{#1}^\infty} % anciennement \overline{#1}

\newcommand\lub{\mathop{\mathrm{lub}}\nolimits}
\newcommand\Lub{\mathop{\mathrm{Lub}}\nolimits}
\newcommand\Ind{\mathop{\mathrm{Ind}}\nolimits}
\newcommand\Max{\mathop{\mathrm{Max}}\nolimits}
\newcommand\Min{\mathop{\mathrm{Min}}\nolimits}
\newcommand\Idl{\mathop{\mathrm{Idl}}\nolimits}
\newcommand\Pfx{\mathop{\mathrm{Pfx}}\nolimits}

\newcommand\send[1]{\mathtt{send}_{#1}}
\newcommand\recv[1]{\mathtt{recv}_{#1}}

\begin{document}

\title[Complete WSTS]{Forward Analysis for WSTS, Part~{II:} Complete WSTS}

\author[A.~Finkel]{Alain Finkel\rsuper a}	%required
\address{{\lsuper a}LSV, ENS Cachan, CNRS, France}	%required
\email{finkel@lsv.ens-cachan.fr}  %optional
\thanks{{\lsuper a}The first author is supported by the french Agence Nationale de la 
Recherche, REACHARD (grant ANR-11-BS02-001)}	%optional

\author[J.~Goubault-Larrecq]{Jean Goubault-Larrecq\rsuper b}	%optional
\address{{\lsuper b}LSV, ENS Cachan, CNRS, France and INRIA Saclay, France}	%optional
\email{goubault@lsv.ens-cachan.fr}  %optional
%\thanks{thanks 2, optional.}	%optional

% \author[]{Author 3}	%optional
% \address{address 3}	%optional
% \email{author3@email3}  %optional
% \thanks{thanks 3, optional.}	%optional

%% etc.

%% required for running head on odd and even pages, use suitable
%% abbreviations in case of long titles and many authors:

%% mandatory lists of keywords and classifications:
\titlecomment{An extended abstract already appeared in Proc. 36th {I}ntl. {C}oll. {A}utomata, {L}anguages and {P}rogramming ({ICALP}'09).}

%%%%%%%%%%%%%%%%%%%%%%%%%%%%%%%%%%%%%%%%%%%%%%%%%%%%%%%%%%%%%%%%%%%%%%%%%%%

%% the abstract has to PRECEED the command \maketitle:
%% be sure not to issue the \maketitle command twice!

\begin{abstract}
  \noindent We describe a simple, conceptual forward analysis
  procedure for $\infty$-complete WSTS $\mathfrak S$.  This computes
  the so-called {\em clover\/} of a state.  When $\mathfrak S$ is the
  completion of a WSTS $\mathfrak X$, the clover in $\mathfrak S$ is a
  finite description of the downward closure of the reachability set.
  We show that such completions are $\infty$-complete exactly when
  $\mathfrak X$ is an {\em $\omega^2$-WSTS\/}, a new robust class of
  WSTS.  We show that our procedure terminates in more cases than the
  generalized Karp-Miller procedure on extensions of Petri nets and on
  lossy channel systems.  We characterize the WSTS where our procedure
  terminates as those that are {\em clover-flattable\/}.  Finally, we
  apply this to well-structured counter systems.
\end{abstract}

\maketitle

%% start the paper here:
\section{Introduction}

\subsection*{Context}

Well-structured transition systems (WSTS) are a general class of
infinite-state systems where coverability---given states $s, t$,
decide whether $s \mathrel{(\geq;\to^*;\geq)} t$, i.e., whether $s
\geq s_1 \to^* t_1 \geq t$ for some $s_1$, $t_1$---is decidable, using
a simple algorithm that works backwards
\cite{F87,F90,FS:wsts,AbdullaCJT00}.

The starting point of this paper and of its first part
\cite{FGL-stacs2009} is our desire to derive similar algorithms
working {\em forwards\/}, namely algorithms computing the {\em
  cover\/} $\dc Post^*(\dc s)$ of $s$.  While the cover allows one to
decide coverability as well, by testing whether $t \in \dc Post^*(\dc
s)$, it can also be used to decide $U$-boundedness, i.e., to decide
whether there are only finitely many states $t$ in the upward-closed
set $U$ and such that $s \mathrel{(\geq;\to^*)} t$.  ($U$-boundedness
generalizes the boundedness problem, which is the instance of
$U$-boundedness where $U$ is the entire set of states).  No backward
algorithm can decide this.  In fact, $U$-boundedness is undecidable in
general, e.g., on lossy channel systems \cite{GC-AF-SPI-IC-96}.  So
the reader should be warned that computing the cover is not possible
for general WSTS.  Despite this, the known forward algorithms are felt
to be more efficient than backward procedures in general: e.g., for
lossy channel systems, although the backward procedure always
terminates, only a (necessarily non-terminating) forward procedure is
implemented in the TREX tool \cite{ABJ:SRE}.  Another argument in
favor of forward procedures is the following: for depth-bounded
processes, a fragment of the $\pi$-calculus, the backward algorithm of
\cite{AbdullaCJT00} is not applicable when the maximal depth of
configurations is not known in advance because, in this case, the
predecessor configurations are not effectively computable
\cite{DBLP:conf/fossacs/WiesZH10}. But the Expand, Enlarge and Check
forward algorithm of \cite{RB07}, which operates on complete WSTS,
solves coverability even though the depth of the process is not known
a priori \cite{DBLP:conf/fossacs/WiesZH10}.

\subsection*{State of the Art}

Karp and Miller \cite{KM:petri} proposed an algorithm, for Petri nets,
which computes a finite representation of the {\em cover\/}, i.e., of
the downward closure of the reachability set of a Petri net.
% $Covering(s_0)$; this algorithm was designed for
%vectors of positive integers (the numerous properties of
%integers hide the abstract reasons for which this algorithm is possible and terminates). 
Finkel \cite{F87,F90} introduced the framework of WSTS and generalized
the Karp-Miller procedure to a class of WSTS. This was achieved by
building a non-effective completion of the set of states, and
replacing $\omega$-accelerations of increasing sequences of states (in
Petri nets) by least upper bounds.  In \cite{Emerson&Namjoshi98,F90} a
variant of this generalization of the Karp-Miller procedure was
studied; but no guarantee was given that the cover could be
represented finitely.  In fact, no effective finite representations of
downward-closed sets were given in \cite{F90}.  Finkel \cite{F91}
modified the Karp-Miller algorithm to reduce the size of the
intermediate computed trees.  Geeraerts {\em et al.\/} \cite{RB07}
recently proposed a weaker acceleration, which avoids some possible
underapproximations in \cite{F91}.  Emerson and Namjoshi
\cite{Emerson&Namjoshi98} take into account the labeling of WSTS and
consequently adapt the generalized Karp-Miller algorithm to
model-checking.  They assume the existence of a compatible dcpo, and
generalize the Karp-Miller procedure to the case of broadcast
protocols (which are equivalent to transfer Petri nets).  However,
termination is then not guaranteed \cite{Esparza&Finkel&Mayr99}, and
in fact neither is the existence of a finite representation of the
cover.  We solved the latter problem in \cite{FGL-stacs2009}.

%%%%%%%%% wsts
Abdulla, Collomb-Annichini, Bouajjani and Jonsson proposed a forward
procedure for lossy channel systems \cite{AbdullaCBJ04} using downward-closed regular languages as symbolic representations. Ganty,
Geeraerts, Raskin and Van Begin \cite{GRvB:eec,RB06} proposed a
forward procedure for solving the coverability problem for WSTS
equipped with an effective adequate domain of limits, or equipped with
a finite set $D$ used as a parameter to tune the precision of an
abstract domain.  Both solutions ensure that every downward-closed set
has a finite representation.  Abdulla {\em et al.\/}
\cite{AbdullaCBJ04} applied this framework to Petri nets and lossy
channel systems. Abdulla, Deneux, Mahata and Nyl\'en proposed a symbolic
framework for dealing with downward-closed sets for Timed Petri nets
\cite{AbdullaDMN04}.

\subsection*{Our Contribution}

First, we define a {\em complete WSTS\/} as a WSTS $\mathfrak S$ whose well-ordering is
also a continuous dcpo (a dcpo is a directed complete partial ordering).  This allows us to design a conceptual
procedure {\bf Clover}$_{\mathfrak S}$ that looks for a finite
representation of the downward closure of the reachability set, i.e.,
of the cover \cite{F90}.  We call such a finite representation a {\em
  clover\/} (for {\em clo\/}sure of {\em cover\/}).  This clearly
separates the fundamental ideas from the data structures used in
implementing Karp-Miller-like algorithms.  Our procedure also
terminates in more cases than the well-known (generalized) Karp-Miller
procedure \cite{Emerson&Namjoshi98,F90}.  We establish the main
properties of clovers in Section~\ref{sec:clover} and use them to
prove {\bf Clover}$_{\mathfrak S}$ correct, notably, in
Section~\ref{sec:KM}.

Second, we characterize complete WSTS for which {\bf
  Clover}$_{\mathfrak S}$ terminates.  These are the ones that have a
(continuous) flattening with the same clover. This establishes a
surprising relationship with the theory of flattening
\cite{bardin-flat-05}.  The result (Theorem~\ref{thm:coverflat}),
together with its corollary on covers, rather than clovers
(Theorem~\ref{thm:cover:flat}), is the main achievement of this paper.

Third, and building on our theory of completions \cite{FGL-stacs2009},
we characterize those WSTS whose completion is a complete WSTS in the
sense above.  They are exactly the {\em $\omega^2$-WSTS\/}, i.e.,
those whose state space is $\omega^2$-wqo (a wqo is a well quasi-ordering), as we show in
Section~\ref{sec:omega2}.  All naturally occurring WSTS are in fact
$\omega^2$-WSTS.  We shall also explain why this study is important:
despite the fact that {\bf Clover}$_{\mathfrak S}$ cannot terminate on
all inputs, that $\mathfrak S$ is an $\omega^2$-WSTS will ensure {\em
  progress\/}, i.e., that every opportunity of accelerating a loop
will eventually be taken by {\bf Clover}$_{\mathfrak S}$.
%!!! mais dans les non clover-flattable?

Finally, we apply our framework of complete WSTS to counter systems in
Section~\ref{sec:counter}. We show that affine counter systems may be
completed into $\infty$-complete WSTS iff the domains of the monotonic
affine functions are upward-closed.

\section{Preliminaries}
\label{sec:prelim}

%\subsection{Complete and Continuous Well-Ordered Sets}

\subsection{Posets, Dcpos}
\label{sec:prelim:dcpo}

We borrow from theories of order, as used in model-checking
\cite{AbdullaCJT00,FS:wsts}, and also from domain theory
\cite{AJ:domains,GHKLMS:contlatt}.  A {\em quasi-ordering\/} $\leq$ is
a reflexive and transitive relation on a set $X$.  It is a (partial)
{\em ordering\/} iff it is antisymmetric.

We write $\geq$ for the converse quasi-ordering, $<$ for the associated strict
ordering (${\leq} \setminus {\geq}$), and $>$ the converse (${\geq}
\setminus {\leq}$) of $<$.  There is also an associated equivalence relation
$\equiv$, defined as ${\leq} \cap {\geq}$.

A set $X$ with a partial ordering $\leq$ is a {\em poset\/} $(X,
\leq)$, or just $X$ when $\leq$ is clear.  If $X$ is merely
quasi-ordered by $\leq$, then the quotient $X / {\equiv}$ is ordered
by the relation induced by $\leq$ on equivalence classes.  So there is
not much difference in dealing with quasi-orderings or partial
orderings, and we shall essentially be concerned with the latter.

% Although
% the consideration of quasi-orderings instead of partial orderings
% helps in the theory of well quasi-orderings, we shall essentially be
% concerned with partial orderings instead.

The {\em upward closure\/} $\upc E$ of a set $E$ in $X$ is $\{y \in X
\mid \exists x \in E \cdot x \leq y\}$.  The {\em downward closure\/}
$\dc E$ is $\{y \in X \mid \exists x \in E \cdot y \leq x\}$.  A
subset $E$ of $X$ is {\em upward-closed\/} if and only if $E = {\upc
  E}$.
% , i.e.,
% any element greater than or equal to some element in $E$ is again in
% $E$.
{\em Downward-closed\/} sets are defined similarly.
%   When the ambient
% space $X$ is not clear from context, we shall write $\dc_X E$, $\upc_X
% E$ instead of $\dc E$, $\uparrow E$.
A {\em basis\/} of a downward-closed (resp.\ upward-closed) set $E$ is
a subset $A$ such that $E=\dc A$ (resp.\ $E = \upc A$); $E$ has a {\em
  finite basis\/} iff $A$ can be chosen to be finite.

A quasi-ordering is {\em well-founded\/} iff it has no infinite
strictly descending chain $x_0 > x_1 > \ldots > x_i > \ldots$\@ An
{\em antichain\/} is a set of pairwise incomparable elements.  A
quasi-ordering is {\em well\/} iff it is well-founded and has no
infinite antichain; equivalently, from any infinite sequence $x_0,
x_1, \ldots, x_i, \ldots$, one can extract an infinite ascending chain
$x_{i_0} \leq x_{i_1} \leq \ldots \leq x_{i_k} \leq \ldots$, with $i_0
< i_1 < \ldots < i_k < \ldots$.  While {\em wqo\/} stands for
well-quasi-ordered set, we abbreviate well posets as {\em wpos\/}.

An {\em upper bound\/} $x \in X$ of $E \subseteq X$ is such that $y
\leq x$ for every $y \in E$.  The \emph{least upper bound (lub)} of a
set $E$, if it exists, is written $\lub(E)$.  An element $x$ of $E$ is
{\em maximal\/} (resp.\ minimal) iff $\upc x \cap E = \{x\}$ (resp.\
$\dc x \cap E = \{x\}$).  Write $\Max E$ (resp.\ $\Min E$) for the set of
maximal (resp.\ minimal) elements of $E$.

A {\em directed subset\/} of $X$ is any non-empty subset $D$ such that
every pair of elements of $D$ has an upper bound in $D$.  Chains,
i.e., totally ordered subsets, and one-element sets are examples of
directed subsets.  A {\em dcpo\/} is a poset in which every directed
subset has a least upper bound.  For any subset $E$ of a dcpo $X$, let
$\Lub (E) = \{\lub (D) \mid D \mbox{ directed subset of } E\}$.
Clearly, $E \subseteq \Lub (E)$; $\Lub (E)$ can be thought of $E$ plus
all limits from elements of $E$.

The {\em way below\/} relation $\ll$ on a dcpo $X$ is defined by $x
\ll y$ iff, for every directed subset $D$ such that $\lub (D) \leq y$,
there is a $z \in D$ such that $x \leq z$.  Note that $x \ll y$
implies $x \leq y$, and that $x' \leq x \ll y \leq y'$ implies $x' \ll
y'$.  Write
% $\uuarrow E = \{y \in X \mid \exists x \in E \cdot x \ll y\}$,
$\ddarrow E = \{y \in X \mid \exists x \in E \cdot y \ll x\}$, and
$\ddarrow x = \ddarrow \{x\}$.  $X$ is {\em continuous\/} iff, for
every $x \in X$, $\ddarrow x$ is a directed subset, and has $x$ as
least upper bound.

When $\leq$ is a well partial ordering that also turns $X$ into a
dcpo, we say that $X$ is a {\em directed complete well order\/}, or
{\em dcwo\/}.  We shall be particularly interested in continuous
dcwos.
%  If additionally $X$ is continuous, we say that $X$ is a
% {\em cdcwo\/}.
% [jgl] Non, ca commence a faire beaucoup d'acronymes, je vire cdcwo.

A subset $U$ of a dcpo $X$ is (Scott-){\em open\/} iff $U$ is
upward-closed, and for any directed subset $D$ of $X$ such that $\lub
(D) \in U$, some element of $D$ is already in $U$.  A map $f : X \to
X$ is (Scott-){\em continuous\/} iff $f$ is monotonic ($x \leq y$
implies $f (x) \leq f (y)$) and for every directed subset $D$ of $X$,
$\lub (f (D)) = f (\lub (D))$.  Equivalently, $f$ is continuous in the
topological sense, i.e., $f^{-1} (U)$ is open for every open $U$.

A weaker requirement is $\omega$-continuity: $f$ is {\em
  $\omega$-continuous\/} iff $\lub \{f (x_n) \mid n \in \nat\} = f (\lub
\{x_n \mid n \in \nat\})$, for every countable chain ${(x_n)}_{n \in
  \nat}$.  This is all we require when we define accelerations, but
general continuity is more natural in proofs.  We won't discuss this
any further: the two notions coincide when $X$ is countable, which
will always be the case of the state spaces $X$ we are interested in,
where the states should be representable on a Turing machine, hence at
most countably many.

The {\em closed\/} sets are the complements of open sets.  Every
closed set is downward-closed.  On a dcpo, the closed subsets are the
subsets $B$ that are both downward-closed and {\em inductive\/}, i.e.,
such that $\Lub (B) = B$.  An inductive subset of $X$ is none other
than a sub-dcpo of $X$.

The {\em closure\/} $cl (A)$ of $A \subseteq X$ is the smallest closed
set containing $A$.  This should not be confused with the {\em
  inductive closure\/} $\Ind (A)$ of $A$, which is obtained as the
smallest inductive subset $B$ containing $A$.  In general, $\dc A
\subseteq \Lub (\dc A) \subseteq \Ind (\dc A) \subseteq cl (A)$, and
all inclusions can be strict.  Consider $X = \nat_\omega^k$, where $k
\in \nat$, and $\nat_\omega$ denotes $\nat$ with a new element
$\omega$ added, ordered by $(n_1, n_2, \ldots, n_k) \leq (n'_1, n'_2,
\ldots, n'_k)$ iff $(n_1, n_2, \ldots, n_k) = (n'_1, n'_2, \ldots,
n'_k)$, or for some $i$, $1\leq i\leq k$,
$n_1=n'_1=n_2=n'_2=\ldots=n_{i-1}=n'_{i-1} = \omega$, $n_i \neq
\omega$, and either $n'_i = \omega$ or $n_i < n'_i \in \nat$.  Then
take $A = \nat^k \subseteq X$: $\dc A = A$, but $\Lub (\dc A) =
\nat_\omega \times \nat^{k-1}$ is strictly larger; in fact $\Lub (\Lub
(\dc A)) = \nat_\omega^2 \times \nat^{k-2}$ is even larger, \ldots,
$\Lub^i (\dc A) = \Lub (\Lub^{i-1} (\dc A))$ equals $\nat_\omega^i
\times \nat^{k-i}$ for all $i$, $2\leq i\leq k$, and this is a
strictly increasing chain of subsets.  All of them are contained in
$Ind (\dc A) = \nat_\omega^k$, which coincides with $cl (A)$ here.  It
may also be the case that $\Ind (\dc A)$ is strictly contained in $cl
(A)$: consider the set $X$ of all pairs $(i,m)$ with $i \in \{0,1\}$,
$m \in \nat$, plus a new element $\omega$, ordered by $(i,m) \leq
(j,n)$ iff $i=j$ and $m=n$, and $(i,m) \leq \omega$ for all $(i,m) \in
X$, and let $A = \{(0, m) \mid m \in \nat\}$; Then $\Ind (\dc A) = A
\cup \{\omega\}$, but the latter is not even downward-closed, so is
strictly smaller than $cl (A)$; in fact $cl (A)$ is the whole of $X$.

All this nitpicking is irrelevant when $X$ is a {\em continuous\/}
dcpo, and $A$ is downward-closed in $X$.  In this case indeed, $\Lub
(A) = \Ind (A) = cl (A)$.  This is well-known, see e.g.,
\cite[Proposition~3.5]{FGL-stacs2009}, and will play an important role
in our constructions. As a matter in fact, the fact that $\Lub (A) =
cl (A)$, in the particular case of continuous dcpos, is required for
lub-accelerations to ever reach the closure of the set of states that
are reachable in a transition system.

\subsection{Well-Structured Transition Systems}

A \emph{transition system} is a pair ${\mathfrak S}=(S,
{\rightarrow})$ of a set $S$, whose elements are called {\em
  states\/}, and a {\em transition relation\/} ${\rightarrow}
\subseteq S \times S$. We write $s \rightarrow s'$ for $(s,s') \in
{\rightarrow}$.  Let $\stackrel{*}{\rightarrow}$ be the transitive and
reflexive closure of the relation $\rightarrow$.  We write
$Post_{{\mathfrak S}}(s)=\{s' \in S \mid s \rightarrow s'\}$ for the
set of immediate successors of the state $s$. The \emph{reachability
  set} of a transition system ${\mathfrak S}=(S, \rightarrow)$ from an
initial state $s_0$ is $Post_{{\mathfrak S}}^*(s_0) = \{s \in S \mid
s_0 \stackrel{*}{\rightarrow} s \}$.
%   The \emph{reachability tree}
% $RT(S,\rightarrow,s_0)$ of a transition system $(S,\rightarrow)$ with
% an initial state $s_0$ is defined as follows: the root is labeled by
% $s_0$ and there is an arc between two nodes $n,n'$ labeled by the
% states $s,s'$ iff $s \rightarrow s'$.

%%The  \emph{reachability graph}
%$RG(S,\rightarrow,s_0)$ of a transition system $(S,\rightarrow)$ with an initial state
%$s_0$ is defined as follows: the set of nodes is the reachability set
%and there is an arc between two states $s,s'$ if  $s \rightarrow s'$.

% In the case of a labeled transition system, we may associate a
% language to a system: the \emph{language} of an $F$-labeled transition
% system is the set of words that label the paths of its reachability
% tree, i.e., $L({\mathfrak S},s_0)=\{ g \in F^* \mid s_0
% \stackrel{g}{\rightarrow} s, s \in S \}$.

We shall be interested in \emph{effective} transition systems.
Intuitively, a transition system $(S, \rightarrow)$ is effective iff
one can compute the set of successors $Post_{{\mathfrak S}}(s)$ of any
state $s$.  We shall take this to imply that $Post_{{\mathfrak S}}(s)$
is finite, and each of its elements is computable, although one could
imagine that $Post_{{\mathfrak S}}(s)$ be described differently, say
as a regular expression.

Formally, one needs to find a representation of the states $s \in S$.
% One sometimes sees requirements that $S$ should be recursive, or r.e.,
% in the literature, but this is not relevant.
A {\em representation map\/} is any surjective map $r : E \to S$ from
some subset $E$ of $\nat$ to $S$.  If $e \in E$ is such that $r (e) =
s$, then one says that $e$ is a {\em code\/} for the state $s$.
% We
%shall never need to put any constraint on $E$, such as requiring that
%$E$ be recursive, or r.e.

An \emph{effective transition system} is a $4$-tuple $(S, \rightarrow,
r, post)$, where $(S, \rightarrow)$ is a transition system, $r : E \to
S$ is a representation map, and $post : E \to \Pfin (E)$ is a
computable map such that, for every code $e$, $r \langle post (e)
\rangle = Post_{{\mathfrak S}}(r (e))$.  We write $r \langle A
\rangle$ the image $\{r (a) \mid a \in A\}$ of the set $A$ by $r$, and
$\Pfin (E)$ is the set of finite subsets of $E$.  A {\em computable
  map\/} from $E$ to $\Pfin (E)$ is by definition a partial recursive
map $post : \nat \to \Pfin (\nat)$ that is defined on all
elements of $E$, and such that $post (e) \in \Pfin (E)$ for all $e \in
E$. 

For reasons of readability, we shall make an abuse of language, and
say that the pair $(S, \rightarrow)$ is itself an effective transition
system in this case, leaving the representation map $r$ and the $post$
function implicit.

An \emph{ordered} transition system is a triple ${\mathfrak
  S}=(S,\rightarrow, \leq)$ where $(S, \rightarrow)$ is a transition
system and $\leq$ is a partial ordering on $S$.  We say that $(S,
\rightarrow, \leq)$ is \emph{effective} if $(S, \rightarrow)$ is
effective and if $\leq$ is decidable.

This is again an abuse of language: formally, an \emph{effective
  ordered} transition system is a $6$-tuple $(S, \rightarrow, \leq, r,
post, \preceq)$ where $(S, \rightarrow, \leq)$ is an ordered
transition system, $(S, \rightarrow, r, post)$ is an effective
transition system, and $\preceq$ is a decidable relation on $E$ such
that $e \preceq e'$ iff $r (e) \leq r (e')$.  By {\em decidable on
  $E$\/}, we mean that $\preceq$ is a partial recursive map from $\nat
\times \nat$ to the set of Booleans, which is defined on $E \times E$
at least.

We say that ${\mathfrak S}=(S,\rightarrow, \leq)$ is \emph{monotonic}
(resp.\ \emph{strictly monotonic}) iff for every $s,s',s_1 \in S$ such
that $s \rightarrow s'$ and $s_1 \geq s$ (resp. $s_1 > s$), there
exists an $s_1' \in S$ such that $s_1 \stackrel{*}{\rightarrow} s_1'$
and $s_1' \geq s'$ (resp. $s_1' > s'$).  ${\mathfrak S}$ is
\emph{strongly monotonic} iff for every $s,s',s_1 \in S$ such that $s
\rightarrow s'$ and $s_1 \geq s$, there exists an $s_1' \in S$ such
that $s_1 \rightarrow s_1'$ and $s_1' \geq s'$.

{\em Finite\/} representations of $Post_{\mathfrak S}^* (s)$, e.g., as
Presburger formulae or finite automata, usually don't exist even for
monotonic transition systems (not even speaking of being computable).
However, the {\em cover\/} $Cover_{\mathfrak S} (s) = \dc
Post^*_{{\mathfrak S}}(\dc s)$ ($= \dc Post^*_{{\mathfrak S}}(s)$ when
$\mathfrak S$ is monotonic) will be much better behaved.  Note that
being able to compute the cover allows one to decide {\em
  coverability\/}: $s \mathrel{(\geq;\to^*;\geq)} t$ iff $t \in
Cover_{\mathfrak S} (s)$.  In most cases we shall encounter, it will
also be decidable whether a finitely represented cover is finite, or
whether it meets a given upward-closed set $U$ in only finitely many
points.  Therefore {\em boundedness\/} (is $Post_{{\mathfrak S}}^*(s)$
finite?) and {\em $U$-boundedness\/} (is $Post_{{\mathfrak S}}^*(s)
\cap U$ finite?)  will be decidable, too.

%Note that, for any monotone transition system ${\mathfrak S}$,
%$Post_{{\mathfrak S}}^*(\dc s_0) \subseteq Cover_{{\mathfrak S}}(s_0)$.
%= \dc Post_{{\mathfrak S}}^*(\dc s_0)$.

An ordered transition system ${\mathfrak S}=(S,\rightarrow,\leq)$ is a
\emph{Well Structured Transition System} ({\em WSTS\/}) iff
${\mathfrak S}$ is monotonic and $(S,\leq)$ is wpo.  This is our object
of study.

% A WSTS ${\mathfrak S}=(S, \rightarrow,\leq)$ has an \emph{effective
%   pred basis} iff, for every state $s \in S$, a finite set $F_s$ such
% that $\upc Pre_{{\mathfrak S}}(\upc s)=\upc F_s$ is computable.  For
% every WSTS ${\mathfrak S}=(S, \rightarrow,\leq)$ with effective pred
% basis, and for every state $s \in S$, the set $Pre_{{\mathfrak
%     S}}^*(\upc s)$ is effectively computable
% \cite{FS:wsts,AbdullaCJT00}.  This entails the decidability of the
% coverability problem.

For strictly monotonic WSTS, it is also possible to decide the
boundedness problem, with the help of the Finite Reachability Tree
(FRT) \cite{F90}. However, the place-boundedness problem (i.e., to
decide whether a place can contain an unbounded number of tokens)
remains undecidable for transfer Petri nets \cite{DufourdFS98}, which
are strictly monotonic WSTS, but it is decidable for Petri nets. It is
decided with the help of a richer structure than the FRT, the
Karp-Miller tree. The set of labels of the Karp-Miller tree is a
finite representation of the cover.

We will consider transition systems that are {\em functional\/}, i.e.,
defined by a finite set of transition functions.  This is, as in
\cite{FGL-stacs2009}, for reasons of simplicity.  However, our {\bf
  Clover}$_{\mathfrak S}$ procedure (Section~\ref{sec:KM}), and
already the technique of {\em accelerating loops\/}
(Definition~\ref{defn:accel}) depends on the considered transition
system being functional.

Formally, a \emph{functional transition system} $(S,
\stackrel{F}{\rightarrow})$ is a labeled transition system where the
transition relation $\stackrel{F}{\rightarrow}$ is defined by a finite
set $F$ of partial functions $f: S \longrightarrow S$, in the sense
that for every $s,s' \in S$, $s \stackrel{F}{\rightarrow} s'$ iff $s'
= f (s)$ for some $f \in F$.  If additionally, a partial ordering
$\leq$ is given, a map $f : S \to S$ is {\em partial monotonic\/} iff
$\dom f$ is upward-closed and for all $x, y \in \dom f$ with $x \leq
y$, $f (x) \leq f (y)$.  An {\em ordered functional transition
  system\/} is an ordered transition system ${\mathfrak
  S}=(S,\stackrel{F}{\rightarrow}, \leq)$ where $F$ consists of
partial monotonic functions.  This is always strongly monotonic.  A
{\em functional WSTS\/} is an ordered functional transition system
where $\leq$ is a well-ordering.

A functional transition system $(S,\stackrel{F}{\rightarrow})$ is
\emph{effective} if every $f \in F$ is computable: given a state $s$
and a function $f$, we can decide whether $s \in \dom f$ and in this
case, one can also compute $f(s)$.

For example, every Petri net, every reset/transfer Petri net, and in
fact every affine counter system (see Definition~\ref{defn:ACS}) is an
effective, functional WSTS.

Lossy channel systems \cite{AbdullaCBJ04} are not functional: any
channel can lose a letter at any position, and although one may think
of encoding this as a functional transition system defined by
functions $f_i$ for each $i$, where $f_i$ would lose the letter at
position $i$, this would require an unbounded number of functions.
However, for the purpose of computing covers, lossy channel systems
are equivalent \cite{Sch-tacs2001} (``equivalent'' means that the
decidability status of the usual properties is the same for both
models) to {\em functional-lossy\/} channel systems, which are
functional \cite{FGL-stacs2009}.  In the latter, there are functions
$\send a$ to add a fixed letter $a$ to the back of each queue (i.e.,
$\dom (\send a) = \Sigma^*$, where $\Sigma$ is the queue alphabet, and
$\send a (w) = wa$), and functions $\recv a$ to read a fixed letter
$a$ from the front of each queue, where reading is only defined when
there is an $a$ in the queue, and means removing all letters up to and
including the first $a$ from the queue (i.e., $\dom(\recv a) = \{w a
w' \mid w, w' \in \Sigma^*\}$, $\recv a (w a w') = w'$ where $a$ does
not occur in $w$).

%%%%%%%%%%%%%%%%%%%%%%%%%%%%%%%%%%%%%%%%%%%%%%%%%%%%%%%%%
%%%%%%%%%%%%%%%%%%%%%%%%%%%%%%%%%%%%%%%%%%%%%%%%%%%%%%%%%
\section{Clovers of Complete WSTS}
\label{sec:clover}
%%%%%%%%%%%%%%%%%%%%%%%%%%%%%%%%%%%%%%%%%%%%%%%%%%%%%%%%%
%%%%%%%%%%%%%%%%%%%%%%%%%%%%%%%%%%%%%%%%%%%%%%%%%%%%%%%%%%%%%%%%%%%%%%%%%%%%%%%%%%%%%%%%%%%%%%%%%%%%%%%%%%%%%%%%%%

\subsection{Complete WSTS and Their Clovers}

All forward procedures for WSTS rest on completing the given WSTS to
one that includes all limits.  E.g., the state space of Petri nets is
$\nat^k$, the set of all markings on $k$ places, but the Karp-Miller
algorithm works on $\nat_\omega^k$, where $\nat_\omega$ is $\nat$ plus
a new top element $\omega$, with the usual componentwise ordering.
We have defined general completions of wpos, serving as state spaces,
and have briefly described completions of (functional) WSTS in
\cite{FGL-stacs2009}.  We temporarily abstract away from this, and
consider {\em complete\/} WSTS directly.

Generalizing the notion of continuity to partial maps, we define:
\begin{defi}
  \label{defn:partial:cont}
  A {\em partial continuous\/} map $f : X \to X$, where $(X, \leq)$ is
  a dcpo, is a partial map whose domain $\dom f$ is open (not just
  upward-closed), and such that for every directed subset $D$ {\em in
    $\dom f$\/}, $\lub (f (D)) = f (\lub (D))$.
\end{defi}
This is the special case of a more topological definition: in general,
a partial continuous map $f : X \to Y$ is a partial map whose domain
is open in $X$, and such that $f^{-1} (U)$ is open (in $X$, or
equivalently here, in $\dom f$) for any open $U$ of $Y$.

The composition of two partial continuous maps again yields a partial
continuous map.
\begin{defi}[Complete WSTS]
  \label{defn:complete:WSTS}
  A {\em complete\/} transition system is a functional transition
  system ${\mathfrak S}=(S,\stackrel{F}{\rightarrow},\leq)$ where $(S,
  \leq)$ is a continuous dcwo and every function in $F$ is partial
  continuous.

  A {\em complete WSTS\/} is a functional WSTS that is complete as a
  functional transition system.
\end{defi}
The point in complete WSTS is that one can {\em accelerate\/} loops:
\begin{defi}[Lub-acceleration]
  \label{defn:accel}
  Let $(X, \leq)$ be a dcpo, $f : X \to X$ be  partial
  %$\omega$-
  continuous.  The {\em lub-acceleration\/} $\accel f : X \to X$ is
  defined by: $\dom \accel f = \dom f$, and for any $x \in \dom f$, if
  $x < f (x)$ then $\accel f (x) = \lub \{f^n (x) \mid n \in \nat\}$,
  else $\accel f (x) = f (x)$.
\end{defi}
Note that if $x \leq f(x)$, then $f (x) \in \dom f$, and $f (x) \leq
f^2 (x)$.  By induction, we can show that $\{f^n (x) \mid n \in
\nat\}$ is an increasing sequence, so that the definition makes sense.

Complete WSTS are strongly monotonic.  One cannot decide, in general,
whether a recursive function $f$ is monotonic \cite{FinkelMP04} or
continuous, whether an ordered set $(S, \leq)$ with a decidable
ordering $\leq$, is a dcpo or whether it is a wpo.  To show the latter
claim for example, fix a finite alphabet $\Sigma$, and consider
subsets $S$ of $\Sigma^*$ specified by a Turing machine $\mathcal M$
with tape alphabet $\Sigma$, so that $S$ is the language accepted by
$\mathcal M$.  Let $\leq$ be, say, the prefix ordering on $\Sigma^*$.
The property that $(S, \leq)$ is a dcpo, resp.\ a wpo, is non-trivial
and extensional, hence undecidable by Rice's Theorem.

We can also prove that given an effective ordered functional
transition system, one cannot decide whether it is a WSTS, or a
complete WSTS, in a similar way.  However, the completion of {\em
  any\/} functional $\omega^2$-WSTS is complete, as we shall see in
Theorem~\ref{thm:omega2}.

\begin{figure}
  \centering
  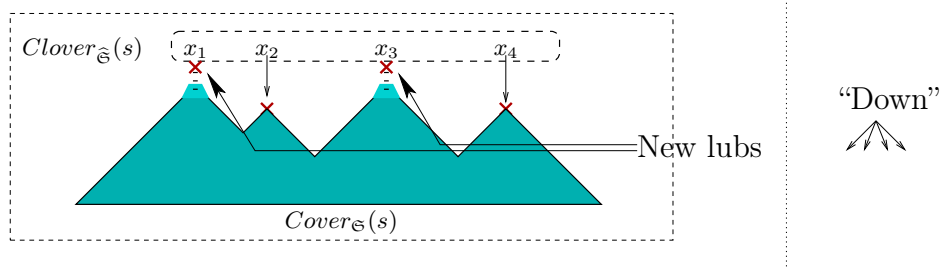
  \caption{The clover and the cover, in a complete space}
  \label{fig:clover1}
\end{figure}

In a complete WSTS, there is a {\em canonical\/} finite representation
of the cover: the {\em clover\/} (a succinct description of the
\emph{cl}osure of the c\emph{over}).
\begin{defi}[Clover]
  \label{defn:clover}
  Let ${\mathfrak S}=(S,\stackrel{F}{\rightarrow},\leq)$ be a complete
  WSTS.  The {\em clover\/} $Clover_{{\mathfrak S}}(s_0)$ of the state
  $s_0 \in S$ is $\Max \Lub(Cover_{{\mathfrak S}}(s_0))$.
\end{defi}
This is illustrated in Figure~\ref{fig:clover1}.  The ``down'' part on
the right is meant to illustrate in which directions one should travel
to go down in the chosen ordering.  The cover $Cover_{{\mathfrak
    S}}(s_0)$ is a downward-closed subset, illustrated in blue (grey
if you read this in black and white).  $\Lub (Cover_{{\mathfrak
    S}}(s_0))$ has some new least upper bounds of directed subsets,
here $x_1$ and $x_3$.  The clover is given by just the maximal points
in $\Lub (Cover_{{\mathfrak S}}(s_0))$, here $x_1$, $x_2$, $x_3$,
$x_4$.

The fact that the clover is indeed a representation of the cover
follows from the following.
\begin{lem}
  \label{lemma:maxlub}
  Let $(S, \leq)$ be a continuous dcwo.  For any closed subset $F$ of
  $S$, $\Max F$ is finite and $F = \dc \Max F$.
\end{lem}
\proof
%We propose two proofs.
%
%The first proof rests on the theory of Noetherian spaces of the second
%author \cite{Gou-lics07}.  Corollary~6.5 of op.cit.\ states that every
%closed subset $F$ of a sober Noetherian space $S$ has the desired
%property.  Since $S$ is a wpo, it is Noetherian in its Scott topology
%(Proposition~3.1 of op.cit.), and since $S$ is a continuous dcpo, $S$
%is also sober \cite[Proposition~7.2.27]{AJ:domains}.
%
%This argument in fact works even outside the realm of continuous wpos,
%and in any sober Noetherian space, but requires some knowledge of
%Noetherian spaces.  

%	Here is a more elementary argument.  
As $F$ is
closed, it is inductive (i.e., $\Lub(F)=F$).  In particular, every element $x$ of $F$ is
below some maximal element of $F$.  This is a well-known, and an easy
application of Zorn's Lemma.  Since $F$ is downward-closed, $F = \dc
\Max F$.  Now every two elements of $\Max F$ are incomparable, i.e.,
$\Max F$ is an antichain: since $S$ is wpo, $\Max F$ is finite.  \qed

\begin{remark}
  Lemma~\ref{lemma:maxlub} generalizes to Noetherian spaces, which
  extend wqos \cite{Gou-lics07}: every closed subset $F$ of a sober
  Noetherian space $S$ is of the form $\dc \Max F$, with $\Max F$
  finite \cite[Corollary~6.5]{Gou-lics07}.  Wpos are sober, and every
  continuous dcpo is sober in its Scott topology
  \cite[Proposition~7.2.27]{AJ:domains}.
%This argument in fact works even outside the realm of continuous wpos,
%and in any sober Noetherian space, but requires some knowledge of
%Noetherian spaces. 
\end{remark}

\begin{prop}
  \label{prop:clover}
  Let ${\mathfrak S}=(S,\stackrel{F}{\rightarrow},\leq)$ be a complete
  WSTS, and $s_0 \in S$.  Then $Clover_{{\mathfrak S}}(s_0)$ is
  finite, and $cl (Cover_{{\mathfrak S}}(s_0)) = \dc
  Clover_{{\mathfrak S}}(s_0)$.
\end{prop}
\proof $\Lub(Cover_{{\mathfrak S}}(s_0)) = cl (Cover_{{\mathfrak
    S}}(s_0))$ since $Cover_{{\mathfrak S}}(s_0)$ is downward-closed,
and $S$ is a continuous dcpo.  Now use Lemma~\ref{lemma:maxlub} on the
closed set $\Lub(Cover_{{\mathfrak S}}(s_0))$.  \qed

For any other representative, i.e., for any finite set $R$ such that
$\dc R= \dc Clover_{{\mathfrak S}}(s_0)$, $Clover_{{\mathfrak
    S}}(s_0)= \Max R$.  Indeed, for any two finite sets $A, B
\subseteq S$ such that $\dc A= \dc B$, $\Max A=\Max B$.  So $Clover$
is the \emph{minimal representative} of the cover, i.e., there is no
representative $R$ with $|R| < |Clover_{{\mathfrak S}}(s_0)|$.  The
clover was called the minimal coverability set in \cite{F91}.
%There is a main problem about the computation of a finite basis of the clover set.  

Despite the fact that the clover is always finite, it is
non-computable in general (see Proposition~\ref{prop:clover:undec}
below).  Nonetheless, it is computable on {\em flat\/} complete WSTS,
and even on the larger class of {\em clover-flattable\/} complete WSTS
(Theorem~\ref{thm:coverflat} below).

\subsection{Completions}

Many WSTS are not complete: the set $\nat^k$ of states of a Petri net
with $k$ places is not even a dcpo.  The set of states of a lossy
channel system with $k$ channels, $(\Sigma^*)^k$, is not a dcpo for
the subword ordering either.  We have defined general completions of
wpos, and of WSTS, in \cite{FGL-stacs2009}, a construction which we
recall quickly.

The {\em completion\/} $\widehat X$ of a wpo $(X, \leq)$ is defined in
any of two equivalent ways.  First, $\widehat X$ is the {\em ideal
  completion\/} $\Idl (X)$ of $X$, i.e., the set of ideals of $X$,
ordered by inclusion, where an {\em ideal\/} is a downward-closed
directed subset of $X$.  The least upper bound of a directed family of
ideals ${(D_i)}_{i \in I}$ is their union.  $\widehat X$ can also be
described as the sobrification $\Sober (X_a)$ of the Noetherian space
$X_a$, but this is probably harder to understand.

There is an embedding $\eta_X : X \to \widehat X$, i.e., an injective
map such that $x \leq x'$ in $X$ iff $\eta_X (x) \leq \eta_X (x')$ in
$\widehat X$.  This is defined by $\eta_X (x) = \dc x$.  This allows
us to consider $X$ as a subset of $\widehat X$, by equating $X$ with
its image $\eta_X \langle X \rangle$, i.e., by equating each element
$x \in X$ with $\dc x \in \widehat X$.  However, we shall only do this
in informal discussions, as this tends to make proofs messier.

For instance, if $X = \nat^k$, e.g., with $k=3$, then $(1,3,2)$ is
equated with the ideal $\dc (1,3,2)$, while $\{(1,m,n) \mid m,n \in
\nat\}$ is a {\em limit\/}, i.e. an element of $\widehat X \setminus
X$; the latter is usually written $(1,\omega,\omega)$, and is the
least upper bound of all $(1,m,n)$, $m, n \in \nat$.  The
downward-closure of $(1,\omega,\omega)$ in $\widehat X$, intersected
with $X$, gives back the set of non-limit elements $\{(1,m,n) \mid m,n
\in \nat\}$.

This is a general situation: one can always write $\widehat X$ as the
disjoint union $X \cup L$, so that any downward-closed subset $D$ of
$X$ can be written as $X \cap \dc A$, where $A$ is a {\em finite\/}
subset of $X \cup L$.  Then $L$, the set of limits, is a {\em weak
  adequate domain of limits\/} (WADL) for $X$---we slightly simplify
Definition~3.1 of \cite{FGL-stacs2009}, itself a slight generalization
of \cite{GRvB:eec}.  In fact, $\widehat X$ (minus $X$) is the {\em
  smallest\/} WADL \cite[Theorem~3.4]{FGL-stacs2009}.

$\widehat X = \Idl (X)$ is always a continuous dcpo.  In fact, it is
even algebraic \cite[Proposition~2.2.22]{AJ:domains}.  It may however
fail to be well, hence to be a continuous dcwo, see
Proposition~\ref{prop:rado:Snotwqo} below.

We have also described a hierarchy of datatypes on which completions
are effective \cite[Section~5]{FGL-stacs2009}.  Notably, $\widehat\nat
= \nat_\omega$, $\widehat A = A$ for any finite poset, and
$\widehat{\prod_{i=1}^k X_i} = \prod_{i=1}^k \widehat X_i$.  Also,
$\widehat {X^*}$ is the space of {\em word-products\/} on $X$.  These
are the products, as defined in \cite{ABJ:SRE}, i.e., regular
expressions that are products of {\em atomic expressions\/} $A^*$ ($A
\in \Pfin (\widehat X)$, $A \neq \emptyset$) or $a^?$ ($a \in \widehat
X$).  In any case, elements of completions $\widehat X$ have a finite
description, and the ordering $\subseteq$ on elements of $\widehat X$
is decidable \cite[Theorem~5.3]{FGL-stacs2009}.

Having defined the completion $\widehat X$ of a wpo $X$, we can define
the completion $\mathfrak S = \widehat{\mathfrak X}$ of a (functional)
WSTS $\mathfrak X = (X,\stackrel{F}{\rightarrow},\leq)$ as $(\widehat
X, \stackrel{\Sober F}{\rightarrow},\subseteq)$, where $\Sober F =
\{\Sober f \mid f \in F\}$ \cite[Section~6]{FGL-stacs2009}.  For each
partial monotonic map $f \in F$, the partial continuous map $\Sober f
: \widehat X \to \widehat X$ is such that $\dom \Sober f = \{C \in
\widehat X \mid C \cap \dom f \neq \emptyset\}$, and $\Sober f (C) =
{\dc f \langle C\rangle}$ for every $C \in \widehat X$.  In the cases
of Petri nets or functional-lossy channel systems, the completed WSTS
is effective \cite[Section~6]{FGL-stacs2009}.

The important fact, which assesses the importance of the clover, is
Proposition~\ref{prop:completion:clover} below.  We first require a
useful lemma.  Up to the identification of $X$ with its image $\eta_X
\langle X \rangle$, this states that for any downward-closed subset
$F$ of $\widehat X$, $cl (F) \cap X = F \cap X$, i.e., taking the
closure of $F$ only adds new limits, no proper elements of $X$.
\begin{lem}
  \label{lemma:cl&X}
  Let $X$ be a wpo.  For any downward-closed subset $F$ of $\widehat
  X$, $\eta_X^{-1} (cl (F)) = \eta_X^{-1} (F)$.
\end{lem}
\proof We show that $\eta_X^{-1} (cl (F)) \subseteq \eta_X^{-1} (F)$;
the converse inclusion is obvious.  Since $\widehat X = \Idl (X)$ is a
continuous dcpo, $cl (F) = \Lub (F)$.  Take any $x \in \eta_X^{-1} (cl
(F))$: then $\eta_X (x) = \dc x$ is the least upper bound of a
directed family of ideals $D_i$ in $F$, $i \in I$: $\dc x = \bigcup_{i
  \in I} D_i$.  So $x$ is in $D_i$ for some $i \in I$, hence $\eta_X
(x) = \dc x \subseteq D_i$, i.e., $\eta_X (x)$ is below $D_i$ in
$\widehat X$.  Since $F$ is downward-closed and $D_i \in F$, $\eta_X
(x)$ is also in $F$, i.e., $x \in \eta_X^{-1} (F)$.  \qed

\begin{figure}
  \centering
  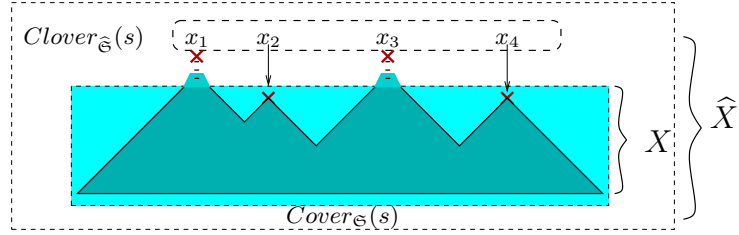
  \caption{The clover and the cover, in a completed space}
  \label{fig:clover}
\end{figure}

Up to the identification of $X$ with $\eta_X \langle X\rangle$, the
next proposition states that $Cover_{\mathfrak X} (s_0) =
Cover_{\mathfrak S} (s_0) \cap X = \dc Clover_{\mathfrak S} (s_0) \cap
X$.  In other words, to compute the cover of $s_0$ in the WSTS
$\mathfrak X$ on the state space $X$, one can equivalently compute the
cover $s_0$ in the completed WSTS $\widehat {\mathfrak X}$, and keep
only those non-limit elements (first equality of
Proposition~\ref{prop:completion:clover}).  Or one can equivalently
compute the {\em closure\/} of the cover in the completed WSTS
$\widehat {\mathfrak X}$, in the form of the downward closure $\dc
Clover_{\mathfrak S} (s_0)$ of its clover.  The closure of the cover
will include extra limit elements, compared to the cover, but no
non-limit element by Lemma~\ref{lemma:cl&X}.  This is illustrated in
Figure~\ref{fig:clover}.

\begin{prop}
  \label{prop:completion:clover}
  Let $\mathfrak S = \widehat{\mathfrak X}$ be the completion of the
  functional WSTS $\mathfrak X = (X, \stackrel{F}{\rightarrow},
  \leq)$.  For every state $s_0 \in X$, $Cover_{\mathfrak X} (s_0) =
  \eta_X^{-1} (Cover_{\mathfrak S} (\eta_X (s_0))) = \eta_X^{-1} (\dc
  {Clover_{\mathfrak S} (\eta_X (s_0))})$.
\end{prop}
\proof The first equality
% of  Proposition \ref{prop:completion:clover}
actually follows from Proposition~6.1 of \cite{FGL-stacs2009}. To be
self-contained, we give a direct proof: this will be a consequence of
(1) and (2) below. The second equality is a consequence of
Proposition~\ref{prop:clover} and Lemma~\ref{lemma:cl&X}.

First, we show that: (1) $\eta_X^{-1} (Cover_{\mathfrak S} (\eta_X
(s_0))) \subseteq Cover_{\mathfrak X} (s_0)$.  Let $x$ be any element
of $\eta_X^{-1} (Cover_{\mathfrak S} (\eta_X (s_0)))$, i.e., $\dc x$
is in $Cover_{\mathfrak S} (\eta_X (s_0))$.  By definition, there is a
natural number $k$, and $k+1$ elements $C_0 = \eta_X (s_0)$, $C_1$,
\ldots, $C_k$ in $\widehat X$, and $k$ partial monotonic maps $f_1$,
\ldots, $f_k$ in $F$ such that $\dc x \subseteq C_k$, and $C_i =
\Sober {f_i} (C_{i-1})$ for every $i$, $1\leq i\leq k$.

Since $\dc x \subseteq C_k = \Sober {f_k} (C_{k-1}) = \dc f_k \langle
C_{k-1}\rangle$, there is an element $x_{k-1} \in C_{k-1} \cap \dom
f_k$ such that $x \leq f_k (x_{k-1})$.  Similarly, there is an
$x_{k-2} \in C_{k-2} \cap \dom f_{k-1}$ such that $x_{k-1} \leq
f_{k-1} (x_{k-2})$, \ldots, an $x_1 \in C_1 \cap \dom f_2$ such that
$x_2 \leq f_2 (x_1)$, and an $x_0 \in C_0 \cap \dom f_1$ such that
$x_1 \leq f_1 (x_0)$.  Since $C_0 = \eta_X (s_0) = \dc s_0$, we have
$x_0 \leq s_0$.  Using the fact that $f_1$, \ldots, $f_k$ are partial
monotonic, $x \leq f_k (f_{k-1} (\ldots (f_2 (f_1 (s_0))))$, so $x \in
Cover_{\mathcal X} (s_0)$.

Conversely, we show: (2) $Cover_{\mathcal X} (s_0) \subseteq
\eta_X^{-1} (Cover_{\mathfrak S} (\eta_X (s_0)))$.  Let $x \in
Cover_{\mathcal X} (s_0)$.  So there is a natural number $k \in \nat$,
and there are $k$ maps $f_1$, \ldots, $f_k$ in $F$ such that $x \leq
f_k (f_{k-1} (\ldots (f_2 (f_1 (s_0)))))$; the latter notation in
particular implies that $f_i (\ldots (f_2 (f_1 (s_0))))$ is defined
for all $i$, $0\leq i\leq k$.  For every $i$, $0 \leq i\leq k$, define
$C_i$ as $\dc f_i (f_{i-1} (\ldots (f_2 (f_1 (s_0)))))$.  We claim
that whenever $i\geq 1$, $C_i = \Sober {f_i} (C_{i-1})$.  Indeed,
$\Sober {f_i} (C_{i-1}) = \dc f_i \langle C_{i-1}\rangle = \dc f_i
\langle\dc f_{i-1} (\ldots (f_2 (f_1 (s_0))))\rangle$.  Since $f_i$ is
partial monotonic, $\dc f_i (\dc y) = \dc f_i (y)$ for every $y$.  So
$\Sober {f_i} (C_{i-1})
  % = \dc f_i (f_{i-1} (\ldots (f_2 (f_1 (s_0)))))
= C_i$.  Next, $C_0 = \dc s_0$; and $\dc x \subseteq C_k$, since $x
\in C_k$ and $C_k$ is downward-closed.  So $\dc x$ is in
$Cover_{\mathcal S} (\dc s_0)$, i.e., $\eta_X (x)$ is in
$Cover_{\mathcal S} (\eta_X (s_0))$.  \qed

$Cover_{\mathfrak S} (s_0)$ is contained, usually strictly, in $\dc
Clover_{\mathfrak S} (s_0)$.  The above states that, when restricted
to non-limit elements (in $X$), both contain the same elements.
Taking lub-accelerations $\accel {(\Sober f)}$ of any composition $f$
of maps in $F$ may leave $Cover_{\mathfrak S} (s_0)$, but is always
contained in $\dc Clover_{\mathfrak S} (s_0) = cl (Cover_{\mathfrak S}
(s_0))$.  So we can safely lub-accelerate in $\mathfrak S =
\widehat{\mathfrak X}$ to compute the clover in $\mathfrak S$.  While
the clover is larger than the cover, taking the intersection back with
$X$ will produce exactly the cover $Cover_{\mathfrak X} (s_0)$.

In more informal terms, the cover is the set of states reachable by
either following the transitions in $F$, or going down.  The closure
of the cover $\dc Clover_{\mathfrak S} (s_0)$ contains not just states
that are reachable in the above sense, but also the limits of chains
of such states.  One may think of the elements of $\dc
Clover_{\mathfrak S} (s_0)$ as being those states that are ``reachable
in infinitely many steps'' from $s_0$.  And we hope to find the
finitely many elements of $Clover_{\mathfrak S} (s_0)$ by doing enough
lub-accelerations.

\section{A Robust Class of WSTS: $\omega^2$-WSTS}
\label{sec:omega2}

It would seem clear that the construction of the completion $\mathfrak
S = \widehat{\mathfrak X}$ of a WSTS $\mathfrak X = (X,
\stackrel{F}{\rightarrow}, \leq)$ be, again, a WSTS.  We shall show
that this is not the case.  The only missing ingredient to show that
$\mathfrak S$ is a complete WSTS is to check that $\widehat X$ is
well-ordered by inclusion.  We have indeed seen that $\widehat X$ is a
continuous dcpo; and $\mathfrak S$ is strongly monotonic, because
$\Sober f$ is continuous, hence monotonic, for every $f \in F$.

Next, we shall concern ourselves with the question: under what
condition on $\mathfrak X$ is $\mathfrak S = \widehat{\mathfrak X}$
again a WSTS?  Equivalently, when is $\widehat X$ well-ordered by
inclusion?  We shall see that there is a definite answer: when $X$ is
$\omega^2$-wqo.

\subsection{Motivation}
\label{sec:omega2:motiv}

The question may seem mostly of academic interest.  Instead, we
illustrate that it is crucial to establish a {\em progress\/} property
described below.

Let us imagine a procedure in the style of the Karp-Miller tree
construction.  We shall provide an abstract version of one, {\bf
  Clover}$_{\mathfrak S}$, in Section~\ref{sec:KM}.  However, to make
things clearer, we shall use a direct imitation of the Karp-Miller
procedure for Petri nets for now, generalized to arbitrary WSTS.  This
is a slight variant of the {\em generalized Karp-Miller procedure\/}
of \cite{F87,F90}, and we shall therefore call it as such.

We build a tree, with nodes labeled by elements of the completion
$\widehat X$, and edges labelled by transitions $f \in F$. During the
procedure, nodes can be marked extensible or non-extensible.  We start
with the tree with only one node labeled $s_0$, and mark it
extensible.  At each step of the procedure, we pick an extensible leaf
node $N$, labeled with $s \in \widehat X$, say, and add new children
to $N$.  For each $f \in F$ such that $s \in \dom \Sober f$, let $s' =
\Sober f (s)$, and add a new child $N'$ to $N$.  The edge from $N$ to
$N'$ is labeled $f$.  If $s'$ already labels some ancestor of $N'$,
then we label $N'$ with $s'$ and mark it non-extensible.  If $s'' \leq
s'$ for no label $s''$ of an ancestor of $N'$, then we label $N'$ with
$s'$ and mark it extensible.  Finally, if $s'' < s'$ for some label
$s''$ of an ancestor $N_0$ of $N'$ (what we shall refer to as case (*)
below), then the path from $N_0$ to $N'$ is labeled with a sequence of
functions $f_1, \ldots, f_p$ from $F$, and we label $N'$ with the
lub-acceleration $\accel {(f_p \circ \ldots \circ f_1)} (s'')$.
(There is a subtle issue here: if there are several such ancestors
$N_0$, then we possibly have to lub-accelerate several sequences $f_1,
\ldots, f_p$ from the label $s''$ of $N_0$: in this case, we must
create several successor nodes $N'$, one for each value of $\accel
{(f_p \circ \ldots \circ f_1)} (s'')$.)  When $X = \nat^k$ and each $f
\in F$ is a Petri net transition, this is the Karp-Miller procedure,
up to the subtle issue just mentioned, which we shall ignore.

Let us recall that the Karp-Miller tree (and also the reachability
tree) is \emph{finitely branching}, since the set $F$ of functions is
finite. This will allow us to use König's Lemma, which states that any
finitely branching, infinite tree has at least one infinite branch.

The reasons why the original Karp-Miller procedure terminates on
(ordinary) Petri nets are two-fold.  First, when $\widehat X =
\nat_\omega^k$, one cannot lub-accelerate more than $k$ times, because
each lub-acceleration introduces a new $\omega$ component to the label
of the produced state, which will not disappear in later node
extensions.  This is specific to Petri nets, and already fails for
reset Petri nets, where $\omega$ components do disappear.

The second reason is of more general applicability: $\widehat X =
\nat_\omega^k$ is wpo, and this implies that along every infinite
branch of the tree thus constructed, case (*) will eventually happen,
and in fact will happen infinitely many times.  Call this {\em
  progress\/}: along any infinite path, one will lub-accelerate
infinitely often.  In the original Karp-Miller procedure for Petri
nets, this will entail termination. 

As we have already announced, for WSTS other than Petri nets,
termination cannot be ensured.  But at least we would like to ensure
progress.  The argument above shows that progress is obtained provided
$\widehat X$ is wpo (or even just wqo).  {\em This\/} is our main
motivation in characterizing those wpos $X$ such that $\widehat X$ is
wpo again.

\begin{figure}
  \centering
  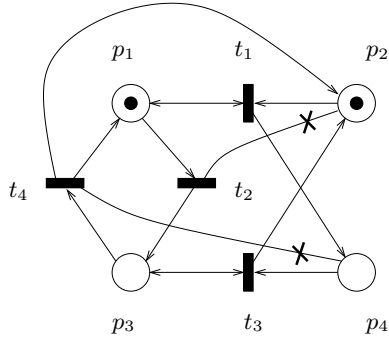
  \caption{The reset Petri net from \cite{DufourdFS98}}
  \label{fig:reset}
\end{figure}
Before we proceed, let us explain why termination cannot be ensured.
Generally, this will follow from undecidability arguments (e.g.,
Proposition~\ref{prop:clover:undec} below). Here is a concrete case of
non-termination. Consider the reset Petri net of
\cite[Example~3]{DufourdFS98}, see Figure~\ref{fig:reset}. This net
has $4$ places and $4$ transitions, hence defines an transition system
on $\nat^4$. Its transitions are: $t_1 (n_1, n_2, n_3, n_4) = (n_1,
n_2-1, n_3, n_4+1)$ if $n_1, n_2 \geq 1$, $t_2 (n_1, n_2, n_3, n_4) =
(n_1-1, 0, n_3+1, n_4)$ if $n_1 \geq 1$, $t_3 (n_1, n_2, n_3, n_4) =
(n_1, n_2+1, n_3, n_4-1)$ if $n_3, n_4 \geq 1$, and $t_4 (n_1, n_2,
n_3, n_4) = (n_1+1, n_2+1, n_3-1, 0)$ if $n_3 \geq 1$. Note that $t_4
(t_3^{n_2} (t_2 (t_1^{n_2} (1, n_2, 0, 0)))) = (1, n_2+1, 0, 0)$
whenever $n_2 \geq 1$.
   % t1 n2 fois -> (1, 0, 0, n_2)
   % t2 -> (0, 0, 1, n_2)
   % t_3 n2 fois -> (0, n_2, 1, 0)
   % t_4 -> (1, n_2+1, 0, 0)
The generalized Karp-Miller tree procedure, starting from $s_0 = (1,
1, 0, 0)$, will produce a child labeled $(1, 0, 0, 1)$ through $t_1$,
then $(0, 0, 1, 1)$ through $t_2$, then $(0, 1, 1, 0)$ through $t_3$.
Using $t_4$ leads us to case (*) with $s' = (1, 2, 0, 0)$.  So the
procedure will lub-accelerate the sequence $t_1 t_2 t_3 t_4$, starting
from $s_0 = (1, 1, 0, 0)$.  However $(t_4 \circ t_3 \circ t_2 \circ
t_1) (s') = (1, 1, 0, 0) = s'$ again, so the sequence of iterates
$(t_4 \circ t_3 \circ t_2 \circ t_1)^n (s_0)$ stabilizes at $s'$, and
$\accel {(t_4 \circ t_3 \circ t_2 \circ t_1)} (s_0) = s'$.  So the
procedure adds a node labeled $s' = (1, 2, 0, 0)$.  Similarly,
starting from the latter, the procedure will eventually lub-accelerate
the sequence $t_1^2 t_2 t_3^2 t_4$, producing a node labeled $(1, 3,
0, 0)$, and in general produce nodes labeled $(1, i+1, 0, 0)$ for any
$i \geq 1$ after having lub-accelerated the sequence
$t_1^it_2t_3^it_4$ from a node labeled $(1, i, 0, 0)$.  In particular,
the generalized Karp-Miller tree procedure will generate infinitely
many nodes, and therefore fail to terminate.

This example also illustrates the following: progress does {\em not\/}
mean that we shall eventually compute limits $\accel g (s)$ that could
not be reached in finitely many steps.  In the example above, we do
lub-accelerate infinitely often, and compute $\accel {(t_4 \circ t_3^i
  \circ t_2 \circ t_1^i)} (1, i, 0, 0)$, but none of these
lub-accelerations actually serve any purpose, since $\accel {(t_4
  \circ t_3^i \circ t_2 \circ t_1^i)} (1, i, 0, 0) = (1, i+1, 0, 0)$
is already equal to $(t_4 \circ t_3^i \circ t_2 \circ t_1^i) (1, i, 0,
0)$.

Progress will take a slightly different form in the actual procedure
{\bf Clover}$_{\mathfrak S}$ of Section~\ref{sec:KM}. In fact, the
latter will not build a tree, as the tree is in fact only algorithmic
support for ensuring a fair choice of a state in $\widehat X$, and
essentially acts as a distraction. However, progress will be crucial
(Proposition ~\ref{prop:clover:stop} states that if the set of values
computed by the procedure {\bf Clover}$_{\mathfrak S}$ is finite then
{\bf Clover}$_{\mathfrak S}$ terminates) in our characterization of
the cases where {\bf Clover}$_{\mathfrak
  S}$ terminates (Theorem~\ref{thm:coverflat}), as those states that
are clover-flattable (see Section~\ref{sec:KM}).  Without it, {\bf
  Clover}$_{\mathfrak S}$ would terminate in strictly less cases.

% We should, by the way, warn the reader that the above imitation of the
% Karp-Miller procedure is in fact {\em not\/} correct in general, in
% the sense that the (intersection of $X$ with) downward-closure of the
% set of states labeling the nodes of the final tree is in general
% strictly smaller than the cover of $s_0$.  As we have said initially,
% this is only meant for illustration.

\subsection{The Rado Structure}
\label{sec:omega2:rado}

We now return to the purpose of this section: showing that $\widehat
X$ is well-ordered iff $X$ is $\omega^2$-wqo.  We start by showing
that , in some cases, $\widehat X$ is indeed {\em not\/} well-ordered.

Take $X$ to be Rado's structure $X_{\text{Rado}}$ \cite{Rado:wqo},
i.e., $\{(m,n) \in \nat^2 \mid m < n\}$, ordered by
$\leq_{\text{Rado}}$: $(m, n) \leq_{\text{Rado}} (m', n')$ iff $m=m'$
and $n\leq n'$, or $n < m'$.  It is well-known that
$\leq_{\text{Rado}}$ is a well quasi-ordering, and that $\pow
(X_{\text{Rado}})$ is not well-quasi-ordered by
$\leq_{\text{Rado}}^\sharp$, defined as $A \leq_{\text{Rado}}^\sharp
B$ iff for every $y \in B$, there is a $x \in A$ such that $x
\leq_{\text{Rado}} y$ \cite{Jancar:wqo:pow}.  (Equivalently, $A
\leq_{\text{Rado}}^\sharp B$ iff $\upc B \subseteq \upc A$.)

Consider indeed $\omega_i = \{(i, n) \mid n \geq i+1\} \cup \{(m,n)
\in X_{\text{Rado}} \mid n \leq i-1\}$, for each $i \in \nat$. This is
pictured as the dark blue (or dark grey) region in
Figure~\ref{fig:rado_D}, and arises naturally in
Lemma~\ref{lemma:Rado:omegai} below. Note that $\omega_i$ is
downward-closed in $\leq_{\text{Rado}}$. Consider the complement
$\overline\omega_i$ of $\omega_i$, and note that $\overline\omega_i
\leq_{\text{Rado}}^\sharp \overline\omega_j$ iff $\upc
{\overline\omega_j} \subseteq \upc {\overline\omega_i}$, iff
$\overline\omega_j \subseteq \overline\omega_i$ (since
$\overline\omega_i$ is upward-closed), iff $\omega_i \subseteq
\omega_j$. However, when $i < j$, $(i,j)$ is in $\omega_i$ but not in
$\omega_j$, so $\overline\omega_i \not\leq_{\text{Rado}}^\sharp
\overline\omega_j$. So ${(\overline\omega_i)}_{i \in \nat}$ is an
infinite sequence of $\pow (X_{\text{Rado}})$ from which one cannot
extract any infinite ascending chain. Hence $\pow (X_{\text{Rado}})$
is indeed not wqo.

Let us characterize $\widehat{X_{\text{Rado}}}$.  To this end, we
exploit the fact that $\widehat{X_{\text{Rado}}} = \Idl
(X_{\text{Rado}})$, and examine the structure of directed subsets of
$X_{\text{Rado}}$.
\begin{lem}
  \label{lemma:Rado:omegai}
  The downward-closed directed subsets of $X_{\text{Rado}}$, apart
  from those of the form $\dc (m,n)$, are of the form $\omega_i =
  \{(i, n) \mid n \geq i+1\} \cup \{(m,n) \in X_{\text{Rado}} \mid n
  \leq i-1\}$, or $\omega=X_{\text{Rado}}$.
\end{lem}
% See Figure~\ref{fig:rado_D} for a pictorial representation of
% $\omega_i$.
\begin{figure}
  \centering
  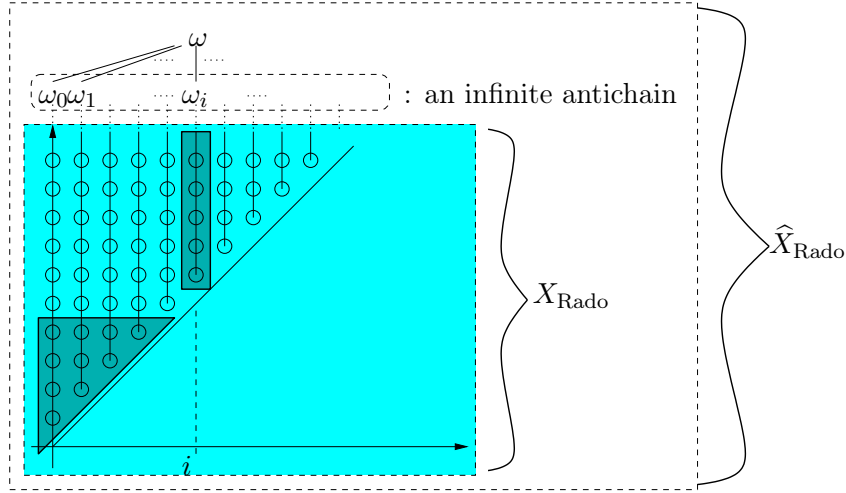
  \caption{Ideals in Rado's Structure}
  \label{fig:rado_D}
\end{figure}

\proof Take any downward-closed directed subset $D$ of
$X_{\text{Rado}}$.  Consider the set $I$ of all integers $i$ such that
some $(i,n)$ is in $D$.  If $I$ is not bounded, then $D =
X_{\text{Rado}}$.  Indeed, for every $(m,n) \in X_{\text{Rado}}$,
since $I$ is not bounded, there is an $(i,n') \in D$ with $i > n$.
Then $(m, n) < (i, n')$, so $(m,n) \in D$.

If $I$ is bounded, on the other hand, let $i$ be the largest element
of $I$.  Then $(i, i+1)$ is in $D$: by assumption $(i,n)$ is in $D$
for some $n \geq i+1$, hence $(i,i+1)$ also, since $D$ is downward-closed.

There cannot be any $(i',j') \in D$ with $i' < i$ and $j' \geq i$.
That is, the rectangular area above the lower triangle of $\omega_i$,
as shown in Figure~\ref{fig:rado_D}, must be entirely outside $D$.
Otherwise, since $D$ is directed, there would be an $(i'', j'') \in D$
with $(i,i+1), (i',j') \leq_{\text{Rado}} (i'', j'')$; the case
$i''=i$ is impossible, since then $(i',j') \leq_{\text{Rado}}
(i'',j'')$ would imply $i'=i''$ and $j' \leq j''$ (impossible since
$i' < i$), or $j' < i''$ (impossible since then $i \leq j' < i'' =
i$); since $i''\neq i$ and $(i,i+1) \leq_{\text{Rado}} (i'', j'')$,
$i'' > i+1$, contradicting the maximality of $i$ in $I$.

On the other hand, since $(i,i+1)$ is in $D$, then the lower triangle
of $\omega_i$, as shown in Figure~\ref{fig:rado_D}, must be in $D$:
these are the points $(m,n)$ with $n<i$.

If the set of natural numbers $n$ such that $(i,n)$ is in $D$ is
bounded, say by $n_{\max}$, then the only elements in $D$ are those of
the form $(i,j)$ with $j \leq n_{\max}$, and those of the form $(m,n)$
with $n<i$.  One checks easily that this is $\dc (i,n_{\max})$ in
$X_{\text{Rado}}$.  Otherwise, $D$ contains every $(i,n)$ with $n \geq
i+1$, and therefore $D$ contains $\omega_i$.  It cannot contain more,
so $D = \omega_i$.  Then one checks that $\omega_i$ is indeed directed
and downward-closed.  \qed

So $\widehat{X_{\text{Rado}}} = \Idl (X_{\text{Rado}})$ is obtained by
adjoining infinitely many elements $\omega_0$, $\omega_1$, \ldots,
$\omega_i$, \ldots, and $\omega$ to $X_{\text{Rado}}$.  They are
ordered so that $(i,n) \leq \omega_i$ for all $n\geq i+1$, $\omega_i
\leq \omega$ for all $i \in \nat$, and no other ordering relationship
exists that involves one of the fresh elements.  In particular, note
that $\{\omega_i \mid i \in \nat\}$ is an infinite antichain, whence
$\widehat{X_{\text{Rado}}} =
%\Sober (X_{Rado\;a}) =
\Idl (X_{\text{Rado}})$ is not wqo:
\begin{prop}
  \label{prop:rado:Snotwqo}
  $\widehat{X_{\text{Rado}}}$ contains an infinite chain, and is
  therefore not well-ordered by inclusion.  \qed
\end{prop}

\subsection{$\omega^2$-WSTS}
\label{sec:omega2:WSTS}

Recall here the working definition in \cite{Jancar:wqo:pow}: a
well-quasi-order $X$ is {\em $\omega^2$-wqo\/} if and only if it does
not contain an (isomorphic copy of) $X_{\text{Rado}}$; here we use
Jan\v car's definition, as it is more tractable than the complex
definition of \cite{MR1219735}.  Jan\v car proved that $X$ is {\em
  $\omega^2$-wqo\/} iff $(\pow (X), \leq^\sharp)$ is wqo, see e.g.\
\cite{Jancar:wqo:pow}.  We show that the above is the only case that
can go bad:
\begin{prop}
  \label{prop:S:wqo}
  Let $S$ be a well-quasi-order.  Then $\widehat S$ is
  well-quasi-ordered by inclusion iff $S$ is $\omega^2$-wqo.
\end{prop}
\proof Recall that $B_1 \leq_{\text{Rado}}^\sharp B_2$ if and only if
for every $y_2 \in B_2$, there is $y_1 \in B_1$ with $y_1
\leq_{\text{Rado}} y_2$.  Note that $B_1 \leq_{\text{Rado}}^\sharp
B_2$ if and only if ${\upc B_1} \supseteq {\upc B_2}$. Reformulate the
previous result of Jan\v car \cite{Jancar:wqo:pow} by using the
ordering $ \leq_{\text{Rado}}^\sharp$: $S$ is $\omega^2$-wqo if and
only if $\pow (S)$ is well-ordered by $\leq_{\text{Rado}}^\sharp$.

Recall that the Alexandroff topology on a poset is the collection of
its upward-closed subsets; i.e., a subset is Alexandroff-open if and
only if it is upward-closed.  Write $S_a$ for $S$ with its Alexandroff
topology.  Any set of the form $\upc B$ in $S$ is Alexandroff-open
(i.e., upward-closed), and any Alexandroff-open is of this form, with
$B$ finite, because $S$ is well.  In other words, the set $\mathcal O
(S_a)$ of all opens (upward-closed subsets) of $S$ is well-ordered by
reverse inclusion $\supseteq$ if and only if $S$ is $\omega^2$-wqo.

Recall that the {\em Hoare powerdomain\/} $\Hoare (S_a)$ of $S_a$ is
the set of all non-empty closed subsets of $S_a$ (the downward-closed
subsets of $S$), ordered by inclusion.  It follows that $\Hoare (S_a)$
is well-ordered by inclusion $\supseteq$ if and only if $S$ is
$\omega^2$-wqo.  Then we recall that $\widehat S = \Sober (S_a)$ is
the subspace of $\Hoare (S)$ consisting of all irreducible closed
subsets \cite{Gou-lics07}.

When $S$ is $\omega^2$-wqo, since $\Hoare (S_a)$ is well-ordered by
inclusion, the smaller set $\widehat S = \Sober (S_a)$ is also
well-ordered by inclusion.

Conversely, assume that $\widehat S = \Sober (S_a)$ is well-ordered by
inclusion.  If $S$ was not $\omega^2$-wqo, then it would contain a
subset $Y$ that is order-isomorphic to $X_{\text{Rado}}$.  Hence
$\widehat S = \Sober (S_a) = \Idl (S)$ would contain $\widehat Y =
\Idl (Y)$.  However by Proposition~\ref{prop:rado:Snotwqo} $\Idl (Y)$
contains an infinite antichain: contradiction.  \qed

Let an {\em $\omega^2$-WSTS\/} be any WSTS whose underlying poset is
$\omega^2$-wqo.  It follows:
\begin{thm}
  \label{thm:omega2}
  Let ${\mathfrak S}=(S,\stackrel{F}{\rightarrow},\leq)$ be a
  functional WSTS.  Then $\widehat{\mathfrak S}$ is a (complete,
  functional) WSTS iff $\mathfrak S$ is an $\omega^2$-WSTS.  \qed
\end{thm}
% Note that $\widehat S = \Idl (S)$ is always an algebraic dcpo
% \cite{AJ:domains}, whence $\widehat S$ is a continuous dcwo
% as soon as $S$ is $\omega^2$-wqo.

\subsection{Are $\omega^2$-wqos Ubiquitous?}
\label{sec:omega2:ubiq}

$X_{\text{Rado}}$ is an example of a wqo that is not $\omega^2$-wqo.
It is natural to ask whether this is the norm or an exception.  We
claim that all wpos used in the verification literature are in fact
$\omega^2$-wpo.

Consider the following grammar of datatypes, which extends
that of \cite[Section~5]{FGL-stacs2009} with the
case of finite trees (last line):
\begin{equation}
  \label{eq:data}
  \begin{array}{rcll}
    D & ::= & \nat & \mbox{natural numbers} \\
    & \mid & A_\leq & \mbox{finite set $A$, ordered by $\leq$}\\
    & \mid & D_1 \times \ldots \times D_k & \mbox{finite product} \\
    & \mid & D_1 + \ldots + D_k & \mbox{finite, disjoint sum} \\
    & \mid & D^* & \mbox{finite words} \\
    & \mid & D^\circledast & \mbox{finite multisets} \\
    & \mid & {\mathcal T} (D) & \mbox{finite trees}
  \end{array}
\end{equation}
$\nat$ is ordered with its usual ordering; the ordering $\leq$ on the
arbitrary finite set $A$ is itself arbitrary.  Finite products are
ordered componentwise: given that each $D_i$ is ordered by $\leq_i$,
then the ordering $\leq$ on $D = D_1 \times \ldots \times D_k$ is
defined by $(x_1, \ldots, x_k) \leq (y_1, \ldots, y_k)$ iff $x_1
\leq_1 y_1$ and \ldots{} and $x_k \leq y_k$.  Finite sums are ordered
in the obvious way: the elements of $D_1 + \ldots + D_k$ are pairs
$(i, x)$ where $1\leq i\leq k$ and $x \in D_i$, and $(i, x) \leq (j,
y)$ iff $i=j$ and $x \leq y$.

$D^*$ is the set of finite words over the (possibly infinite) alphabet
$D$, and given that the ordering on $D$ is $\leq$, $D^*$ is ordered by
the {\em divisibility ordering\/} $\leq^*$, defined by $w \leq^* w'$
iff, writing $w$ as the sequence of letters $a_1 a_2 \ldots a_n$, then
$w'$ is of the form $w_0 a'_1 w_1 \allowbreak a'_2 \ldots \allowbreak
a'_n w_n$, for some words $w_0$, $w_1$, \ldots, $w_n$, and some
letters $a'_i$, $1\leq i\leq n$, such that $a_i \leq a'_i$.

$D^\circledast$ is the set of finite multisets $\mopen x_1, \ldots,
x_n \mclose$ of elements of $D$.  Write again $\leq$ the ordering on
$D$.  Then $D^\circledast$ is ordered by $\leq^\circledast$ defined
as: $\mopen x_1, \allowbreak x_2, \ldots, \allowbreak x_m \mclose
\leq^\circledast \mopen y_1, y_2, \ldots, y_n\mclose$ iff there is an
injective map $r : \{1, \ldots, \allowbreak m\} \to \{1, \ldots,
\allowbreak n\}$ such that $x_i \leq y_{r (i)}$ for all $i$, $1\leq
i\leq m$.

Note that $\leq^\circledast$ is not the usual multiset extension
$\leq^{mul}$ of $\leq$.  However, for one, this is the $\leq^m$
quasi-ordering considered, on finite sets $X$, by Abdulla {\em et
  al.\/} \cite[Section~2]{DBLP:conf/formats/AbdullaDMN04} for example.
Then, it turns out that $m \leq^\circledast m'$ entails $m \leq^{mul}
m'$.  In particular, the fact that $\leq^\circledast$ is well,
whenever $\leq$ is, {\em entails\/} that $\leq^{mul}$ is well: given
any sequence of multisets ${(m_i)}_{i \in \nat}$, one can extract an
infinite ascending chain with respect to $\leq^\circledast$, hence
also with respect to $\leq^{mul}$.  Similarly, when $(D^\circledast,
\leq^\circledast)$ is an $\omega^2$-wqo, then so is $(D^\circledast,
\leq^{mul})$, using the fact that $X$ is $\omega^2$-wqo iff both $X$
and $\pow (X)$ are wqo (the latter, equipped with $\leq^\sharp$).

Finally, ${\mathcal T} (D)$ is the set of all finite (unranked,
ordered) trees over function symbols taken from $D$.  This is the
smallest set $X$ such that, for every $f \in D$, for every $\vec t \in
D^*$, the pair $(f, \vec t)$ is in $X$.  When $\vec t$ is the word
consisting of the terms $t_1 t_2 \ldots t_m$, we usually write $(f,
\vec t)$ as the {\em term\/} $f (t_1, t_2, \ldots, t_m)$.  Given an
ordering $\leq$ on $D$, the {\em embedding ordering\/} $\leq^{emb}$ on
${\mathcal T} (D)$ is defined by induction on the sum of the sizes of
the terms to compare by: $t = f (t_1, t_2, \ldots, t_m) \leq^{emb} g
(u_1, u_2, \ldots, u_n)$ iff $t \leq^{emb} u_j$ for some $j$, $1\leq
j\leq n$, or $f \leq g$ and $t_1 t_2 \ldots t_m
\mathrel{{(\leq^{emb})}^*} u_1 u_2 \ldots u_n$.

We will prove that every datatype defined in (\ref{eq:data}) is not
only $\omega^2$-wqo but a better quasi-ordering (bqo).  Better quasi-orderings were
invented by Nash-Williams to overcome certain limitations of wqo
theory \cite{NashWilliams:inftree}.  Their definition is complex, and
we shall omit it.  For short, $X$ is bqo iff $\pow^{\omega_1} (X)$ is
wqo, where $\omega_1$ is the first uncountable ordinal, $\pow^\alpha
(X)$ is defined for every ordinal $\alpha$ by $\pow^0 (X) = X$,
$\pow^{\alpha+1} = \pow (\pow^\alpha (X))$, $\pow^\alpha (X) =
\bigcup_{\beta < \alpha} \pow^\beta (X)$ for every limit ordinal
$\alpha$, and where powersets are quasi-ordered by $\leq^\sharp$.
Abdulla and Nyl\'en give a gentle introduction to the theory of bqos
\cite{AN:bqo}.

Then:
\begin{prop}
  \label{prop:data:bqo}
  Every datatype defined in (\ref{eq:data}) is $\omega^2$-wqo, and in
  fact bqo.
\end{prop}
% refs trouvees dans sterken-kissig-introduction-to-wqo-and-bqo-theory.pdf, mais pas verifiees!
\proof 
%Better quasi-orderings (bqo) were invented by Nash-Williams to
%overcome certain limitations of wqo theory
%\cite{NashWilliams:inftree}.  Their definition is complex, and we
%shall omit it.  For short, $X$ is wqo iff $\pow^{\omega_1} (X)$ is wqo,
%where $\omega_1$ is the first uncountable ordinal,
%$\pow^\alpha (X)$ is defined for every ordinal $\alpha$ by $\pow^0 (X) = X$,
%$\pow^{\alpha+1} = \pow (\pow^\alpha (X))$, $\pow^\alpha (X) =
%\bigcup_{\beta < \alpha} \pow^\beta (X)$ for every limit ordinal
%$\alpha$, and where powersets are quasi-ordered by $\leq^\sharp$.
%Abdulla and Nyl\'en give a gentle introduction to the theory of bqos
%\cite{AN:bqo}.

Every bqo is $\omega^2$-wqo, as the above characterization shows
($\pow^\alpha (X)$ is wqo for all $\alpha \leq \omega_1$, hence
certainly for $\alpha=0$ and $\alpha=1$).  Any finite ordered set, any
finite union of bqos, any finite product of bqos is bqo
\cite{Milner:wqo}.  When $D$ is bqo, the set of all ordinal-indexed
sequences over $D$ is again bqo under an obvious extension of the
divisibility ordering, see \cite{NashWilliams:inftree} or
\cite[2.22]{Milner:wqo}.  Since any subset of a bqo is again bqo,
we deduce that $D^*$ is bqo whenever $D$ is (this is also mentioned in
\cite[Theorem~3.1 (3)]{AN:bqo}).  When $D$ is bqo, $D^\circledast$ is proved to be a
bqo in \cite[Theorem~3.1 (4)]{AN:bqo}.  Finally, $D$ is bqo implies
that ${\mathcal T} (D)$ is bqo by \cite[Theorem~2.2]{Laver:Fraisse};
Laver in fact shows that the class of so-called $Q$-trees is bqo under
tree embedding as soon as $Q$ is, where a $Q$-tree is a possibly
infinitely branching tree with branches of length at most $\omega$
whose nodes are labeled with elements of $Q$.  \qed

In fact, all naturally occurring wqos are bqos, perhaps to the notable
exception of finite graphs quasi-ordered by the graph minor relation,
which are wqo \cite{RS:minors:XX} but not known to be bqo.
% cf. forster-bqos-and-induction.pdf

\subsection{Effective Complete WSTS}
\label{sec:omega2:eff}

The completion $\widehat{\mathfrak S}$ of a WSTS $\mathfrak S$ is
effective iff the completion $\widehat S$ of the set of states is
effective and $\Sober f$ is recursive for all $f \in F$.  $\widehat S$
is effective for all the data types of
\cite[Section~5]{FGL-stacs2009}\footnote{That is, of
  Section~\ref{sec:omega2:ubiq} of this paper, see (\ref{eq:data}), to
  the exception of the finite tree constructor.  We have a proof that
  $\widehat S$ is in fact effective for all the data types of
  (\ref{eq:data}) \cite{FGL:partI}, but this is not published yet.}.
Also, $\Sober f$ is indeed recursive for all $f \in F$, whether in
Petri nets, functional-lossy channel systems, and reset/transfer Petri
nets notably.

In the case of ordinary or reset/transfer Petri nets, and in general
for all affine counter systems (which we shall investigate from
Definition~\ref{defn:ACS} on), $\Sober f$ coincides with the extension
$\overline f$ defined in \cite[Section~2]{FinkelMP04}: whenever $\dom
f$ is upward-closed and $f : \nat^k \to \nat^k$ is defined by $f (\vec
s) = A \vec s + \vec a$, for some matrix $A \in \nat^{k \times k}$ and
vector $\vec a \in \Z^k$, then $\dom\Sober f = \upc_S {\dom f}$, and
$\Sober (f) (\vec s)$ is again defined as $A \vec s + \vec a$, this
time for all $\vec s\in \nat_\omega^k$, and using the convention that
$0 \times \omega = 0$ when computing the matrix product $A \vec s$
\cite[Theorem~7.9]{FinkelMP04}.

In the case of functional-lossy channel systems, it is easy to see
that $\dom \Sober (\send a) = \widehat S$, $\Sober (\send a) (P) = P
a^?$ for every word-product $P$; and that $\dom \Sober (\recv a) =
\upc_S a^?$, and:
\begin{eqnarray*}
  \Sober (\recv a) (a^? P) & = & P \\
  \Sober (\recv a) (b^? P) & = & \Sober (\recv a) (P) \qquad(b\neq a) \\
  \Sober (\recv a) (A^* P) & = & A^* P \quad\text{if }a \in A \\
  \Sober (\recv a) (A^* P) & = & \Sober (\recv a) (P) \quad\text{otherwise}
\end{eqnarray*}
These formulae in fact work whenever letters are taken from an
alphabet that is wqo; for example, any of the data types $D$ of
(\ref{eq:data}).  We retrieve the formulae of \cite[Lemma~6]{ABJ:SRE},
which were proved in the case where the alphabet $D$ is finite, with
$=$ as ordering.  This also generalizes the algorithms on the
so-called word language generators of \cite{AbdullaDMN04}, which are
elements of $(A^\circledast)^*$ with $A$ finite.

As promised, we can now show:
\begin{prop}
  \label{prop:clover:undec}
  There are effective complete WSTS ${\mathfrak S}$ such that the map
  $Clover_{\mathfrak S} : S \to \Pfin (S)$ is not recursive.
%   One can
%   in particular pick a reset Petri net or a lossy channel system for
%   ${\mathfrak S}$.
\end{prop}
\proof Let $\mathfrak S$ be the completion of a functional-lossy
channel system \cite[Section~6]{FGL-stacs2009} on the message alphabet
$\Sigma$.  By Theorem~\ref{thm:omega2}, $\mathfrak S$ is a complete
WSTS.  It is effective, too, see above or \cite[Lemma~6]{ABJ:SRE}.
$Clover_{\mathfrak S} (s_0)$ can be written as a finite set of tuples,
consisting of control states $q_i$ (one for each of the communicating
automata) and of word-products $P_j$ (one for each channel).  Each
$P_j$ is a product of atomic expressions $A^*$ ($A \in \Pfin
(\Sigma)$, $A \not\emptyset$) or $a^?$ ($a \in \Sigma$).  Now
$Post_{{\mathfrak S}}^*(s_0)$ is finite iff none of these atomic
expressions is of the form $A^*$.  So, if we could compute
$Clover_{\mathfrak S} (s_0)$, this would allow us to decide
boundedness for functional-lossy channel systems.  However
functional-lossy channel systems are equivalent to lossy channel
systems in this respect, and boundedness is undecidable for the latter
\cite{GC-AF-SPI-IC-96}.  We could have played the same argument with
reset Petri nets \cite{DufourdFS98} instead as well.  \qed

\section{A Conceptual Karp-Miller Procedure}
% [jgl] ``Procedure'' was ``procedure'' (capitalization, referee 2)
\label{sec:KM}
%%%%%%%%%%%%%%%%%%%%%%%%%%%%%%%%%%%%%%%%%%%%%%%%%%%%%%%%%
%%%%%%%%%%%%%%%%%%%%%%%%%%%%%%%%%%%%%%%%%%%%%%%%%%%%%%%%%

%{\em Finite\/} representations of $Post_{\mathfrak S}^* (s)$, e.g., as
%Presburger formulae or finite automata, usually don't exist even for
%monotonic transition systems (not even speaking of being computable).
%However, the {\em cover\/} $Cover_{\mathfrak S} (s) = \dc
%Post^*_{{\mathfrak S}}(\dc s)$ ($= \dc Post^*_{{\mathfrak S}}(s)$ when
%$\mathfrak S$ is monotonic) will be much better behaved.  

There are some advantages in using a forward procedure to compute
(part of) the clover for solving coverability. For depth-bounded
processes, a fragment of the $\pi$-calculus, the simple algorithm that
works backward (computing the set of predecessors of an upward-closed
initial set) of \cite{AbdullaCJT00} is not applicable when the maximal
depth of configurations is not known in advance because, in this case,
the predecessor configurations are not effectively computable
\cite{DBLP:conf/fossacs/WiesZH10}. It has been also proved that,
unlike backward algorithms (which solve coverability without computing
the clover), the Expand, Enlarge and Check forward algorithm of
\cite{RB07}, which operates on complete WSTS, solves coverability by
computing a \emph{sufficient} part of the clover, even though the
depth of the process is not known a priori
\cite{DBLP:conf/fossacs/WiesZH10}.
%; so the coverability problem is solved by this way.
Recently, Zufferey, Wies and Henzinger proposed to compute a part of
the clover by using a particular widening, called a \emph{set-widening
  operator} \cite{DBLP:conf/vmcai/ZuffereyWH12}, which loses some
information, but always terminates and seems sufficiently precise to
compute the clover in various case studies.
%  These examples show one of
% the advantages of using a forward procedure in computing (part of) the
% clover for solving coverability.

The Petri net case also gives complexity-theoretic insights.  Solving
coverability in Petri nets can be done by using Rackoff's forward
procedure \cite{DBLP:journals/tcs/Rackoff78}, or the backward
procedure \cite{Bozzelli:2011:CAB:2045343.2045353}.  Both work in
EXPSPACE---the complexity of the forward coverability procedure of
\cite{RB07} is not known.  On the other hand, the complexity of
computing the clover is not primitive recursive for Petri nets
\cite{DBLP:journals/jacm/MayrM81}.

Model-checking safety properties of WSTS can be reduced to
coverability, but there are other properties, such as {\em
  boundedness\/} (is $Post_{{\mathfrak S}}^*(s)$ finite?) and
\emph{$U$-boundedness} (is $Post_{{\mathfrak S}}^*(s) \cap U$ finite?)
that cannot be reduced to coverability: $U$-boundedness is decidable
for Petri nets and for Vector Addition Systems but undecidable for
Reset Vector Addition Systems \cite{DufourdFS98}, and for Lossy
Channel Systems \cite{Mayr:undec}, hence for general WSTS.

Recall that being able to compute the clover allows one to decide not
only {\em coverability\/} since $s \mathrel{(\geq;\to^*;\geq)} t$ iff
$t \in Cover_{\mathfrak S} (s)$ iff $\exists t' \in Clover_{\mathfrak
  S} (s)$ such that $t \leq t'$ but also boundedness, $U$-boundedness
and place-boundedness. To the best of our knowledge, the only known
algorithms that decide place-boundedness (and also some formal
language properties such as regularity and context-freeness of Petri
net languages) \emph{require one to compute the clover}.

Another argument in favor of computing clovers is Emerson and
Namjoshi's \cite{Emerson&Namjoshi98} approach to model-checking
\emph{liveness} properties of WSTS, which uses a finite (coverability)
graph based on the clover. Since WSTS enjoy the finite path property
(\cite{Emerson&Namjoshi98}, Definition~7), model-checking liveness
properties is decidable for complete WSTS for which the clover is
computable.

% %%%%%%%%%%%%%%%%%%%%%%%%%%%%%%%%%%%%%%%%%%%%%%%%%%%%%%%%%%%%%%%%%%%%%%%
%using
%a simple algorithm that works backwards (in computing the set of predecessors of an upward-closed initial set)
%\cite{AbdullaCJT:00,Finkel&Schnoebelen:01} or by using the forward Expand, Enlarge and Check algorithm \cite{GRB06}. The Model Cheking of Safety properties of WSTS can be reduced to coverability but there are other properties, like the $U$-boundedness (to decide whether there are only finitely many states $t$ in the
%upward-closed set $U$ and such that $s \to^* t$) which cannot be reduced to coverability ($U$-boundedness is decidable for Vector Addition Systems (VAS) but it is undecidable for Reset Vector Addition Systems \cite{Dufourd:98}, and for Lossy Channel Systems \cite{Mayr:03}, hence for general WSTS). 
%In general, deciding the $U$-boundedness needs the computation of a finite representation (called the clover in \cite{FGL:II}) of  the downward closure of the reachability set (called the cover). Moreover, following the approach of Emerson and Namjoshi \cite{Emerson:98}, the Model Checking of Liveness properties of WSTS, uses a finite (coverability) graph based on the clover. Since WSTS enjoy the finite path property (\cite{Emerson:98}, Definition 7), the Model Checking of Liveness properties is decidable for complete WSTS for which the clover is computable.

All these reasons motivate us to \emph{try} to compute the clover for
classes of complete WSTS, even though it is not computable in general.

%%%%%%%%%%%%%%%%%%%%%%%%%%%%%%%%%%%%%%%%%%%%%%%%%%%%%%%%%%%%%%%%%%%%%%%
The key to designing some form of a Karp-Miller procedure, such as the
generalized Karp-Miller tree procedure
(Section~\ref{sec:omega2:motiv}) or the {\bf Clover}$_{\mathfrak S}$
procedure below is being able to {\em compute\/} lub-accelerations.
Hence:
\begin{defi}[$\infty$-Effective]
  \label{defn:infty:eff}
  An effective complete functional WSTS ${\mathfrak
    S}=(S,\stackrel{F}{\rightarrow},\leq)$ is
  \emph{$\infty$-effective} iff every function $\accel g$ is
  computable, for every $g \in F^*$, where $F^*$ is the set of all
  compositions of maps in $F$.
\end{defi}
E.g., the completion of a Petri net is $\infty$-effective: not only is
$\nat_\omega^k$ a wpo, but every composition of transitions $g \in
F^*$ is of the form $g (\vec x) = \vec x + \delta$, where $\delta \in
\Z^k$.  If $\vec x < g (\vec x)$ then $\delta \in \nat^k \setminus
\{0\}$.  Write $\vec x_i$ the $i$th component of $\vec x$, it follows
that $\accel g (\vec x)$ is the tuple whose $i$th component is $\vec
x_i$ if $\delta_i=0$, $\omega$ otherwise.

Let $\mathfrak S$ be an $\infty$-effective WSTS, and write $A
\leq^\flat B$ iff $\dc A \subseteq \dc B$, i.e., iff every element of
$A$ is below some element of $B$.  This is the {\em Hoare
  quasi-ordering\/}, also known as the {\em domination\/}
quasi-ordering.  The following is a simple procedure which computes
the clover of its input $s_0 \in S$ (when it terminates):
\begin{center}
  \begin{tabular}{l}
    {\bf Procedure Clover}$_{\mathfrak S} (s_0):$\\
    % {\bf Parameter:} an effective complete WSTS ${\mathfrak S}$; \\
    % {\bf Input:} an initial state $s_0$; \\
    % {\bf Output:} $A$; %$Clover_{{\mathfrak S}}(s_0)$;
    1. $A \leftarrow \{s_0\}$;\\
    2. {\bf while} $Post_{{\mathfrak S}}(A) \not\leq^\flat A$ {\bf do}\\
    \quad (a) Choose fairly (see below) $(g,a) \in F^* \times A$  such that $a
    \in \dom g$; % and $g (a) \not\leq^\flat A$;
    \\
    \quad (b) $A  \leftarrow A \cup \{\accel g(a)\}$;\\
    % \quad (c) $A  \leftarrow \Max A$;\\
    3. {\bf return} $\Max A$;
  \end{tabular}
\end{center}

Note that {\bf Clover}$_{\mathfrak S}$ is well-defined and all its
lines are computable by assumption, provided we make clear what we
mean by fair choice in line (a).  Call $A_m$ the value of $A$ at the
start of the $(m-1)$st turn of the loop at step~2 (so in particular
$A_0 = \{s_0\}$).  The choice at line (a) is {\em fair\/} iff, on
every infinite execution, every pair $(g, a) \in F^* \times A_m$ will
be picked at some later stage $n \geq m$.

A possible implementation of this fair choice is the generalized
Karp-Miller tree construction of Section~\ref{sec:omega2:motiv}:
organize the states of $A$ as labeling nodes of a tree that we grow.
At step $m$, $A_m$ is the set of leaves of the tree, and case (*) of
the generalized Karp-Miller tree construction ensures that all pairs
$(g, a) \in F^* \times A_m$ will eventually be picked for
consideration.  However, the generalized Karp-Miller tree construction
does some useless work, e.g., when two nodes of the tree bear the same
label.

%   First, it is \emph{well defined} because the lub of any
% infinite non-decreasing sequence of states which can be written as
% $(g^i(x))_{i \geq 0}$ exists (because $S$ is a dcpo) and all the
% functions in $F^*$ are $\omega$-continuous.  Every instruction of the
% procedure {\bf Clover}$_{\mathfrak S}$ is \emph{computable}. From the fact that the
% functions $f \in F$ are computable we deduce that every function $g
% \in F^*$ is also computable; and because the complete WSTS is
% effective, all the lub-accelerations $\accel g$ are also computable.
% The $\Max$ operation at line (c) is not necessary, but it allows to
% reduce optimally the size of $A$. We may still optimize in different
% ways this procedure: for instance, one may test whether it is
% necessary to add $\accel g(a)$ by replacing line $(b)$ by: {\bf if}
% $\accel g(a) \not\leq^\flat A$ {\bf then} $A \leftarrow A \cup
% \{\accel g(a)\}$.

Most existing proposals for generalizing the Karp-Miller construction
do build such a tree \cite{KM:petri,F90,F91,RB07}, or a graph
\cite{Emerson&Namjoshi98}.  We claim that this is mere algorithmic
support for ensuring fairness, and that the goal of such procedures is
to compute a finite representation of the cover.  Our {\bf
  Clover}$_{\mathfrak S}$ procedure computes the clover, which is the
minimal such representation, and isolates algorithmic details from the
core construction.

We shall also see that termination of {\bf Clover}$_{\mathfrak S}$ has
strong ties with the theory of \emph{flattening}
\cite{bardin-flat-05}.  However, Bardin {\em et al.\/} require one to
enumerate sets of the form $g^* (\vec x)$, which is sometimes harder
than computing the single element $\accel g (\vec x)$.  For example,
if $g : \nat^k \to \nat^k$ is an affine map $g(\vec x)=A\vec x+\vec
b-\vec a$ for some matrix $A \in \nat^{k \times k}$ and vectors $\vec
a, \vec b\in \nat^k$, then $\accel g(\vec x)$ is computable as a
vector in $\nat_\omega^k$, as we have seen in
Section~\ref{sec:omega2:eff}.  But $g^*(\vec x)$ is not even definable
by a Presburger formula in general, in fact even when $g$ is a
composition of Petri net transitions; this is because reachability
sets of Petri nets are not semi-linear in general \cite{HP:VASS}.
% such that $A^*=\{A^n; n \geq 0 \}$ of $A$
%%%%%%%%%%%%%
%The idea to accelerate circuits can be found in the well-known Karp-Miller algorithm for Petri nets
%\cite{KM:petri} which has been generalized to a class of well-structured transition
%systems \cite{F90}. More precisely, it consists to replace the beginning of an infinite
%repetitive path in the current temporary Generalized Karp and Miller tree in construction $s_1
%\stackrel{g}\rightarrow_{GKM} s_2 ...\stackrel{g}\rightarrow_{GKM}
%s_i...$ with $s_1 \leq s_2 \leq ...\leq s_i  \leq ...$ and $g$ a sequence of transitions by  $s_1
%\stackrel{g}\rightarrow_{GKM} \lub(g^i(s_1)_i)$ (or by a single state equal to
%$\lub(g^i(s_1)_i)$). The acceleration of circuits consists in enumerating all the circuits in the original system and to
%compute the reflexive and
%transitive closure of their reachability sets; 
%hence this strategy computes the set $\{g^i(s); i \geq 0 \}$ if $g$
%labels a circuit and $s$ is a state such that $s \stackrel{g}\rightarrow$.
%The tool FAST for counter systems applies
%this strategy with success \cite{bardin-flat-05}. 
%Here, the difference is that we don't compute
%the exact acceleration of a circuit $g$ but only its lub in the particular
%case where moreover $g(s) > s$. 
%To explore the set of all circuits is the kernel of the theory of
%acceleration and of flattenings which have been presented and used
%mainly for counter systems.
%%%%%%%%

Finally, we use a \emph{fixpoint test} (line~2) that is not in the
Karp-Miller algorithm; and this improvement allows {\bf
  Clover}$_{\mathfrak S}$ to terminate in \emph{more cases} than the
Karp-Miller procedure when it is used for extended Petri nets (for
reset Petri nets for instance, which are a special case of the affine
maps above), as we shall see. To decide whether the current set $A$,
which is always an under-approximation of $Clover_{\mathfrak S}
(s_0)$, is the clover, it is enough to decide whether $Post_{\mathfrak
  S} (A) \leq^\flat A$.  The various Karp-Miller procedures only test
each branch of a tree separately, to the partial exception of the
minimal coverability tree algorithm \cite{F90} and Geeraerts {\em et
  al.\/}'s recent coverability algorithm \cite{RB07}, which compare
nodes across branches.  That the simple test $Post_{\mathfrak S} (A)
\leq^\flat A$ does all this at once does not seem to have been
observed until now.

%%%%%%%%%%%%%%%%%%%%%%%%%%%%%%%%%%%%%%%%%%%%%%%%%%%%%%%%%%
\subsection{Correctness and Termination of the Clover Procedure}
%%%%%%%%%%%%%%%%%%%%%%%%%%%%%%%%%%%%%%%%%%%%%%%%%%%%%%%%%%

% Let us remark that the clover procedure can be defined for any
% $\infty$-effective complete transition system.
By Proposition~\ref{prop:clover:undec}, we cannot hope to have {\bf
  Clover}$_{\mathfrak S}$ terminate on all inputs.  But we can at
least start by showing that it is correct, whenever it terminates.
This will be Theorem~\ref{cor} below.

We first show that if {\bf Clover}$_{\mathfrak S}$ terminates then the
computed set $A$ is contained in $\Lub (Post^*_{{\mathfrak S}}(s_0))$.
It is crucial that $\Lub (F) = cl (F)$ for any downward-closed set
$F$, which holds because the state space $S$ is a continuous dcpo.  We
use this through invocations to Proposition~\ref{prop:clover}.

\begin{lem}
  \label{lemma:Post*:cl}
  Let ${\mathfrak S} = (S, \stackrel{F}{\rightarrow}, \leq)$ be a
  complete (functional) WSTS.  For any subset $A$ of states,
  $Post_{\mathfrak S}^* (cl (A)) \subseteq cl (Post_{\mathfrak S}^*
  (A))$.
\end{lem}
\proof We first observe that $Post_{\mathfrak S} (cl (A)) \subseteq cl
(Post_{\mathfrak S} (A))$.  Indeed, for any $s \in Post_{\mathfrak S}
(cl (A))$, there is an $f \in F$ and some $t \in \dom f \cap cl (A)$
such that $f (t) = s$.  Let $U$ be the complement of $cl
(Post_{\mathfrak S} (A))$: $U$ is open by definition.  Since $f$ is
partial continuous, $f^{-1} (U)$ is open.  If $s$ were in $U$, then
$t$ would be in $f^{-1} (U)$, and in $cl (A)$.  It is a general
property of topological spaces that an open (here $f^{-1} (U)$) meets
$cl (A)$ iff it meets $A$.  So there is also a state $t'$ in $f^{-1}
(U) \cap A$.  That is, $t' \in \dom f$, $f (t') \in U$ and $t' \in A$.
But $t' \in A$ implies $f (t') \in Post_{\mathfrak S} (A) \subseteq cl
(Post_{\mathfrak S} (A))$, contradicting the fact that $f (t') \in U$.
So $s$ cannot be in $U$, i.e., $s \in cl (Post_{\mathfrak S} (A))$.

By an easy induction on $k \in \nat$, it follows that $Post_{\mathfrak
  S}^k (cl (A)) \subseteq cl (Post_{\mathfrak S}^k (A))$, hence that
$Post_{\mathfrak S}^* (cl (A)) \subseteq cl (Post_{\mathfrak S}^*
(A))$.  \qed

\begin{prop}
  \label{prop:An:clover}
  Let ${\mathfrak S}$ be an $\infty$-effective complete functional
  transition system and $A_n$ be the value of the set $A$, computed by
  the procedure {\bf Clover}$_{\mathfrak S}$ on input $s_0$, after $n$
  iterations of the while statement at line $2$.  Then $A_n$ is
  finite, and $A_n \leq^\flat A_{n+1} \leq^\flat Clover_{{\mathfrak
      S}}(s_0)$, for every $n \in \nat$.
\end{prop}
\proof
%   Let $A_n$ be the value of the set $A$, at line $2$, after $n$
%   iterations of the while statement and at the beginning of the $n+1$
%   iteration.
It is obvious that $A_n$ is finite.  Also, the inclusion $A_n
\subseteq \dc A_{n+1}$ is clear, and entails $A_n \leq^\flat A_{n+1}$.

We show that $A_n \leq^\flat Clover_{{\mathfrak S}}(s_0)$, i.e., that
$A_n \subseteq \dc Clover_{{\mathfrak S}}(s_0)$, by induction on $n$.
By Proposition~\ref{prop:clover}, it is equivalent to show that $A_n
\subseteq cl (Cover_{{\mathfrak S}}(s_0))$.

If $n=0, A_0=\{s_0\}$, so $A_0 \subseteq Cover_{{\mathfrak S}}(s_0)
\subseteq cl (Cover_{{\mathfrak S}}(s_0))$.

Assume $A_n \subseteq cl( Cover_{{\mathfrak S}}(s_0))$, and let us
prove that $A_{n+1} \subseteq cl (Cover_{{\mathfrak S}}(s_0))$.  Let
$(g,a)$ be the selected pair at line (a).  We must show that $\accel g
(a) \in cl (Cover_{{\mathfrak S}}(s_0))$.

If $a \not< g (a)$, then $\accel g (a) = g (a)$ is in $Post_{\mathfrak
  S}^* (a)$, and since $a \in A_n$ and $A_n \subseteq cl
(Cover_{{\mathfrak S}}(s_0))$ by induction hypothesis, $g (a)$ is in
$Post_{\mathfrak S}^* (cl (Cover_{{\mathfrak S}}(s_0)))$.  The latter
is contained in $cl ( Post_{\mathfrak S}^* (Cover_{{\mathfrak
    S}}(s_0)))$ by Lemma~\ref{lemma:Post*:cl}, i.e., in $cl
(Cover_{{\mathfrak S}}(s_0))$ by monotonicity.

If $a < g (a)$, then $\accel g (a) = \lub \{g^n (a) \mid n \in \nat\}$
is a least upper bound of a directed chain of elements in
$Post_{\mathfrak S}^* (a)$.  So $\accel g (a) \in \Lub
(Post_{\mathfrak S}^* (a)) \subseteq cl (Post_{\mathfrak S}^* (a))$.
Since $a \in A_n$ and $A_n \subseteq cl (Cover_{{\mathfrak S}}(s_0))$
by induction hypothesis, $\accel g (a)$ is in $cl (Post_{\mathfrak
  S}^* (cl (Cover_{{\mathfrak S}}(s_0))))$.  The latter is contained
in $cl (cl ( Post_{\mathfrak S}^* (Cover_{{\mathfrak S}}(s_0)))) = cl
( Post_{\mathfrak S}^* (Cover_{{\mathfrak S}}(s_0)))$ by
Lemma~\ref{lemma:Post*:cl}, i.e., in $cl (Cover_{{\mathfrak S}}(s_0))$
by monotonicity.  \qed

If the procedure {\bf Clover}$_{\mathfrak S}$ does not stop, it will
compute an infinite sequence of sets of states.  In other words, {\bf
  Clover}$_{\mathfrak S}$ does not deadlock.  This is the progress
property mentioned in Section~\ref{sec:omega2:motiv}.

\begin{prop}[Progress]
  \label{prop:clover:stop}
  Let ${\mathfrak S}$ be an $\infty$-effective complete functional
  WSTS and $A_n$ be the value of the set $A$, computed by the
  procedure {\bf Clover}$_{\mathfrak S}$ on input $s_0$, after $n$
  iterations of the while statement at line $2$.  If $\bigcup_n A_n$
  is finite, then the procedure {\bf Clover}$_{\mathfrak S}$
  terminates on input $s_0$.
\end{prop}
\proof Assume {\bf Clover}$_{\mathfrak S}$ does not stop on input
$s_0$, but $A = \bigcup_n A_n$ is finite.  Since $A_n \leq^\flat
A_{n+1}$, there is an index $m$ such that $A_n=A_m$ for all $n \geq
m$; also $A = A_m$.  Let $(g, a) \in F^* \times A$ be arbitrary.  We
shall show that $g (a) \leq^\flat A$, i.e., there is an element $a'
\in A$ such that $g (a) \leq a'$.  Since $a \in A_m$,
  %either $g (a) \leq^\flat A_m$, and we are done, or
by fairness there is an $n \in \nat$ with $n \geq m$ such that
  %$A_n$ is defined, there is an $a' \in A_n$ such that $a \leq a'$ and
  %either $g (a') \leq^\flat A_n$ or $(g, a')$
$(g, a)$ is picked at line (a) after $n$ iterations of the loop.
%   Since $A_n = A_m = A$, there is an $a' \in A$ such that $a \leq a'$
%   and $g (a') \leq^\flat A$, or $(g, a')$ is picked at line (a).  In
%   the first case, $g (a) \leq g (a') \leq^\flat A$, and we are done.
%  In the second case, $\accel g (a') \leq^\flat A_{n+1} = A$, so $g
%   (a) \leq g (a') \leq \accel g (a') \leq^\flat A$.  Since in all
%   cases $g (a) \leq^\flat A$, $Post_{\mathcal S}^* (A) \leq^\flat
%   A$, hence the procedure must stop after $m$ turns of the loop:
%   contradiction.  So $A$ is infinite.  The converse implication is
%   obvious.  \qed
Then $\accel g (a) \leq^\flat A_{n+1} = A$, so $g (a) \leq \accel g
(a) \leq^\flat A_{n+1} = A$.  It follows that $Post_{\mathcal S}^* (A)
\leq^\flat A$, so $Post_{\mathcal S} (A) \leq^\flat A$, hence the
procedure must stop after $m$ turns of the loop: contradiction.  The
converse implication is obvious.  \qed

While {\bf Clover}$_{\mathfrak S}$ is non-deterministic, this is {\em
  don't care non-determinism\/}: if one execution does not terminate,
then no execution terminates. If {\bf Clover}$_{\mathfrak S}$
terminates, then it computes the clover, and if it does not terminate,
then at each step $n$, the set $A_n$ is contained
in the clover.  Let us recall that $A_n \leq^\flat A_{n+1}$.
We can now prove:

\begin{thm}[Correctness]
  \label{cor}
% [jgl] L'ancienne formulation etait:
%   ``{\bf Clover}$_{\mathfrak S} (s_0)$ terminates iff it computes $Clover_{\mathfrak S} (s_0)$.''
% Le referee 3 se demande ce que ca veut dire.
% En fait, moi non plus je ne trouvais pas ca clair.
% Je remplace par:
  If {\bf Clover}$_{\mathfrak S} (s_0)$ terminates, then it computes
  $Clover_{\mathfrak S} (s_0)$.
\end{thm}
\proof If {\bf Clover}$_{\mathfrak S}$ terminates, then it returns a
set $\Max A$ such that $Post_{{\mathfrak S}}(A) \leq^\flat A$, i.e.,
$Post_{{\mathfrak S}}(A) \subseteq \dc A$.  By monotonicity, it
follows that $Post_{{\mathfrak S}}(\dc A) \subseteq \dc A$, hence that
$\dc {Post_{{\mathfrak S}}(\dc A)} \subseteq \dc A$.  Note that $\dc A
= \dc {\Max A}$, since $A$ is finite.  It follows that
$Cover_{\mathfrak S} (s)$ is contained in $\dc {\Max A}$ for any $s
\in \dc A$.

However, by Proposition~\ref{prop:An:clover}, $\{s_0\} = A_0
\leq^\flat A_1 \leq^\flat \ldots \leq^\flat A_n \leq^\flat \ldots
\leq^\flat A$, so $s_0 \in \dc A$.  So $Cover_{\mathfrak S} (s_0)
\subseteq \dc {\Max A}$.

Since $A$ is finite, $\Max A$ is, too, so $\dc {\Max A}$ is closed.
Any closed set containing another set must contain its closure.  So
$\dc {\Max A}$ must also contain $cl (Cover_{\mathfrak S} (s_0))$.  By
Proposition~\ref{prop:clover}, $\dc {\Max A}$ must therefore contain
$\dc Clover_{\mathfrak S} (s_0)$.  In other words, $Clover_{\mathfrak
  S} (s_0) \leq^\flat \Max A$.  However, using
Proposition~\ref{prop:An:clover} again, $\Max A \leq^\flat A
\leq^\flat Clover_{\mathfrak S} (s_0)$.  So $\Max A =
Clover_{\mathfrak S} (s_0)$.  \qed

%We may improve the previous clover procedure by observing that we may reduce
%the size of the set $A$ in different ways: 

%\begin{enumerate}
%\item We may add, in the procedure clover set {\bf Clover}$_{\mathfrak S}$, at the beginning of
%  the line $3$ the following operation: $A \leftarrow \Max(A)$ and $A$
%  becomes the minimal basis of Clover. Or we could add, at the end
%  of lines $(b)$ and $(c)$ in {\bf Clover}$_{\mathfrak S}$, the operation: $A \leftarrow
%  \Max(A)$ which reduces optimally the size of $A$.
%\item We may also choose to add a new state $\accel g(a)$ in $A$ only if it is
%incomparable with (all the states of) $A$, and in this case,
%immediately after to make a $\Max$-operation on the value of $A$ (see the procedure {\bf Clover}$_{\mathfrak S}$).
%\end{enumerate}

%When the procedure {\bf Clover}$_{\mathfrak S}$ terminates, the size of the set $A$ can be
%non-primitive recursive.
If the generalized Karp-Miller tree procedure (see
Section~\ref{sec:omega2:motiv}) terminates then it has found a finite
set $g_1,g_2,...,g_n$ of maps to lub-accelerate.
%, and the set of labels has the same downward closure as the clover.
These lub-accelerations will also be found by {\bf Clover}$_{\mathfrak
  S}$, by fairness.  From the fixpoint test, {\bf Clover}$_{\mathfrak
  S}$ will also stop.  So {\bf Clover}$_{\mathfrak S}$ terminates on
at least all inputs where the generalized Karp-Miller tree procedure
terminates.  We can say more:
%  We show that it terminates on strictly more inputs:

\begin{prop}
  \label{moreKM}
  The procedure {\bf Clover}$_{\mathfrak S}$ terminates on strictly
  more input states $s_0 \in S$ than the generalized Karp-Miller tree
  procedure.
\end{prop}
\proof Consider the reset Petri net of \cite[Example~3]{DufourdFS98}
again (Figure~\ref{fig:reset}).  Add a new transition $t_5 (n_1, n_2,
n_3, n_4) = (n_1+1, n_2+1, n_3+1, n_4+1)$.  The generalized
Karp-Miller procedure does not terminate on this modified reset Petri
net starting from $s_0 = (1,1,0,0)$, because it already does not
terminate on the smaller one of Section~\ref{sec:omega2:motiv}.  On
the other hand, by fairness, {\bf Clover}$_{\mathfrak S}$ will sooner
or later decide to pick a pair of the form $(t_5, a)$ at line (a), and
then immediately terminate with the maximal state
$(\omega,\omega,\omega,\omega)$, which is the sole element of the
clover.  \qed

Deciding when {\bf Clover}$_{\mathfrak S}$ terminates is itself
impossible.  We first observe that {\bf Clover}$_{\mathfrak S}$
terminates on each bounded state.
\begin{lem}
  \label{lemma:bounded:term}
  Let $\mathfrak S = (S, \stackrel F \rightarrow)$ be an
  $\infty$-effective complete WSTS, and $s_0 \in S$ a state that is
  {\em bounded\/}, i.e., such that the reachability set
  $Post^*_{{\mathfrak S}} (s_0)$ is finite.  Then {\bf
    Clover}$_{\mathfrak S} (s_0)$ terminates.
\end{lem}
\proof Since $Post^*_{{\mathfrak S}} (s_0)$ is finite, $\accel g (s)$
is in $Post^*_{{\mathfrak S}} (s_0)$ for every $s \in
Post^*_{{\mathfrak S}} (s_0)$ and every $g \in F^*$ with $s \in \dom
g$.  So, defining again $A_n$ as the value of the set $A$ computed by
{\bf Clover}$_{\mathfrak S}$ on input $s_0$, after $n$ iterations of
the while statement at line $2$, $\bigcup_{n \in \nat} A_n$ is
contained in $Post^*_{{\mathfrak S}} (s_0)$, hence finite.  By
Proposition~\ref{prop:clover:stop}, {\bf Clover}$_{\mathfrak S} (s_0)$
terminates.  \qed

\begin{prop}
  \label{terminaison-indec}
  There is an $\infty$-effective complete WSTS such that we cannot
  decide, given $s_0 \in S$, whether {\bf Clover}$_{\mathfrak S}
  (s_0)$ will terminate.
\end{prop}
\proof Assume we can decide whether {\bf Clover}$_{\mathfrak S} (s_0)$
terminates.  

If {\bf Clover}$_{\mathfrak S} (s_0)$ does not terminate, then
$Post^*_{{\mathfrak S}} (s_0)$ is infinite, by
Lemma~\ref{lemma:bounded:term}.

If on the other hand {\bf Clover}$_{\mathfrak S} (s_0)$ terminates,
then it computes the clover $Clover_{\mathfrak S} (s_0)$ by
Theorem~\ref{cor}, and we can decide boundedness as in the proof of
Proposition~\ref{prop:clover:undec}, in the case of functional-lossy
channel systems: just check whether any of the computed word-products
contains a starred atomic expression $A^*$.

In any case, we can decide boundedness, i.e., whether
$Post^*_{{\mathfrak S}} (s_0)$ is finite.  But this is impossible
\cite{GC-AF-SPI-IC-96,DBLP:journals/tcs/Mayr03}.  A similar argument
works with reset Petri nets, where boundedness is also undecidable
\cite{DufourdFS98}.  \qed

%%%%%%%%%%%%%%%%%%%%%%%%%%%%%%%%%%%%%%%%%%%%%%%%%%%%%%%%%
\subsection{Clover-Flattable Complete WSTS}
%%%%%%%%%%%%%%%%%%%%%%%%%%%%%%%%%%%%%%%%%%%%%%%%%%%%%%%%%

We now characterize those $\infty$-effective complete WSTS on which
{\bf Clover}$_{\mathfrak S}$ terminates.

% \begin{floatingfigure}{0.2\linewidth}
%   \input{flattening.pdf_t}
%   \caption{Flattening}
%   \label{fig:flattening}
% \end{floatingfigure}

A functional transition system $({\mathfrak
  S},\stackrel{F}{\rightarrow})$ with initial state $s_0$ is
\emph{flat} iff there are finitely many words $w_1,w_2,...,w_k \in
F^*$ such that any fireable sequence of transitions from $s_0$ is
contained in the language $w_1^*w_2^*...w_k^*$.  (We equate functions
in $F$ with letters from the alphabet $F$.)  corresponding composition
of maps, i.e., $fg$ denotes $g \circ f$.)  Ginsburg and Spanier
\cite{GS:bounded} call this a {\em bounded\/} language, and show that
it is decidable whether any context-free language is flat.

\begin{figure}
  \centering
  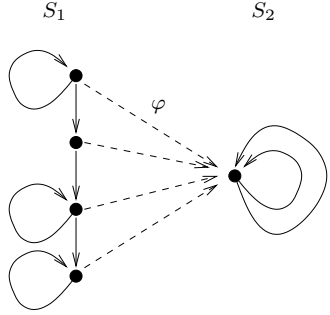 % 70%
  \caption{Flattening}
  \label{fig:flattening}
\end{figure}

Not all systems of interest are flat.  The simplest example of a
non-flat system has one state $q$ and two transitions $q {\buildrel a
  \over\to} q$ and $q {\buildrel b \over\to} q$.

For an arbitrary system $S$, \emph{flattening} \cite{bardin-flat-05}
consists in finding a flat system $S'$, equivalent to $S$ with respect
to reachability, and in computing on $S'$ instead of $S$.  We adapt
the definition in \cite{bardin-flat-05} to functional transition
systems, without an explicit finite control graph for now (but see
Definition~\ref{defn:clover:flattable:strong}).
\begin{defi}[Flattening]
  \label{defn:flatten}
  A \emph{flattening} of a functional transition system ${\mathfrak
    S}_2=(S_2,\stackrel{F_2}{\rightarrow})$ is a pair $({\mathfrak
    S_1}, \varphi)$, where:
  \begin{enumerate}[(1)]
  \item ${\mathfrak S}_1= (S_1, \stackrel{F_1}{\rightarrow})$ is a
    {\em flat\/} functional transition system;
  \item and $\varphi : {\mathfrak S}_1 \to {\mathfrak S}_2$ is a {\em
      morphism\/} of transition systems.  That is, $\varphi$ is a pair
    of two maps, both written $\varphi$, from $S_1$ to $S_2$ and from
    $F_1$ to $F_2$, such that for all $(s,s') \in S_1^2$, for all $f_1
    \in F_1$ such that $s\in \dom f_1$ and $s' = f_1 (s)$, $\varphi
    (s) \in \dom \varphi (f_1)$ and $\varphi (s') = \varphi (f_1)
    (\varphi (s))$ (see Figure~\ref{fig:flattening}).
  \end{enumerate}
\end{defi}

\noindent Let us recall that a pair $({\mathfrak S},s_0)$ of a transition system
and a state is \emph{$Post^*$-flattable} iff there is a flattening
${\mathfrak S}_1$ of ${\mathfrak S}$ and a state $s_1$ of $\mathfrak
S_1$ such that $\varphi (s_1) = s_0$ and $Post_{{\mathfrak S}}^*(s_0)=
\varphi (Post_{{\mathfrak S}_1}^*(s_1))$.

Recall that we equate ordered functional transition systems $(S,
\stackrel F \rightarrow, \leq)$ with their underlying function
transition system $(S, \stackrel F \rightarrow)$.  The notion of
flattening then extends to ordered functional transition systems.
However, it is then natural to consider {\em monotonic flattenings\/},
where in addition $\varphi : S_1 \to S_2$ is monotonic.  In the case
of complete transition systems, the natural extension requires
$\varphi$ to be continuous:
\begin{defi}[Continuous Flattening]
  \label{defn:flatten:cont}
  Let ${\mathfrak S}_2=(S_2,\stackrel{F_2}{\rightarrow}, \leq_2)$ be a
  complete transition system.  A flattening $({\mathfrak S_1},
  \varphi)$ of ${\mathfrak S}_2$ is {\em continuous\/} iff:
  \begin{enumerate}[(1)]
  \item ${\mathfrak S}_1= (S_1, \stackrel{F_1}{\rightarrow}, \leq_1)$ is a
    {\em complete\/} transition system;
  \item and $\varphi : S_1 \to S_2$ is {\em continuous\/}.
  \end{enumerate}
\end{defi}

\begin{defi}[Clover-Flattable]
  \label{defn:clover:flattable}
  Let $\mathfrak S$ be a complete transition system, and $s_0$ be a
  state.  We say that $({\mathfrak S},s_0)$ is \emph{clover-flattable}
  iff there is an continuous flattening $({\mathfrak S}_1, \varphi)$
  of ${\mathfrak S}$, and a state $s_1$ of $\mathfrak S_1$ such that:
  \begin{enumerate}[(1)]
  \item $\varphi (s_1) = s_0$ ($\varphi$ maps initial states to
    initial states);
  \item $cl (Cover_{\mathfrak S} (s_0)) = cl (\varphi \langle cl
    (Cover_{{\mathfrak S}_1} (s_1))\rangle)$ ($\varphi$ preserves the
    closures of the covers of the initial states).
  \end{enumerate}
\end{defi}

\noindent On complete WSTS---our object of study---, the second condition can be
simplified to $\dc {Clover_{{\mathfrak S}} (s_0)} = \dc {\varphi
  (Clover_{{\mathfrak S}_1} (s_1))}$ (using
Proposition~\ref{prop:clover} and the fact that $\varphi$, as a
continuous map, is monotonic), or equivalently to $Clover_{{\mathfrak
    S}} (s_0) = \Max \varphi \langle Clover_{{\mathfrak S}_1} (s_1)
\rangle$.  Recall also that, when ${\mathfrak S}$ is the completion
$\widehat {\mathfrak X}$ of a WSTS ${\mathfrak X} =
(X,\stackrel{F}{\rightarrow},\leq)$, the clover of $s_0 \in X$ is a
finite description of the {\em cover\/} of $s_0$ in $\mathfrak X$
(Proposition~\ref{prop:completion:clover}), and this is what $\varphi$
should preserve, up to taking downward closures.

There are apparently weaker and stronger froms of clover-flattability,
which we now introduce.  Let us start with the weak form, where
equality in the second condition is replaced by inclusion:
\begin{defi}[Weakly Clover-Flattable]
  \label{defn:clover:flattable:weak}
  Let $\mathfrak S$ be a complete transition system, and $s_0$ be a
  state.  We say that $({\mathfrak S},s_0)$ is \emph{weakly
    clover-flattable} iff there is an continuous flattening
  $({\mathfrak S}_1, \varphi)$ of ${\mathfrak S}$, and a state $s_1$
  of $\mathfrak S_1$ such that:
  \begin{enumerate}[(1)]
  \item $\varphi (s_1) \leq s_0$;
  \item and $cl (Cover_{\mathfrak S} (s_0)) \subseteq cl (\varphi
    \langle cl (Cover_{{\mathfrak S}_1} (s_1))\rangle)$.
  \end{enumerate}
\end{defi}

\noindent One may simplify the second condition slightly, to: $Cover_{\mathfrak
  S} (s_0) \subseteq cl (\varphi \langle cl (Cover_{{\mathfrak S}_1}
(s_1))\rangle)$.  In the case of complete WSTS, this is equivalent to
$Clover_{{\mathfrak S}} (s_0) \leq^\flat \varphi (Clover_{{\mathfrak
    S}_1} (s_1))$.

The strong form of clover-flattability uses an explicit finite control
graph, as in \cite{bardin-flat-05}.  Recall that a {\em rlre\/}
(restricted linear regular expression) over the alphabet $\Sigma$ is a
regular expression of the form $w_1^* w_2^*...w_k^*$, where
$w_1,w_2,...,w_k \in \Sigma^*$.  The language of an rlre is clearly
bounded, and the language $\Pfx (w_1^* w_2^* \ldots w_k^*)$ of prefixes
of all words from the latter is then again bounded \cite{GS:bounded}.

Recall that a deterministic finite automaton (DFA) is a tuple
${\mathcal A} = (\Sigma, Q, \delta, q_0, Fin)$, where $\Sigma$ is a
finite alphabet, $Q$ is a finite set of so-called {\em control
  states\/}, $q_0 \in Q$ is the {\em initial\/} state, $Fin \subseteq
Q$ is the set of {\em final\/} states, and $\delta : Q \times \Sigma
\to Q$ is a partial function called the {\em transition function\/}.

\begin{figure}
  \begin{center}
    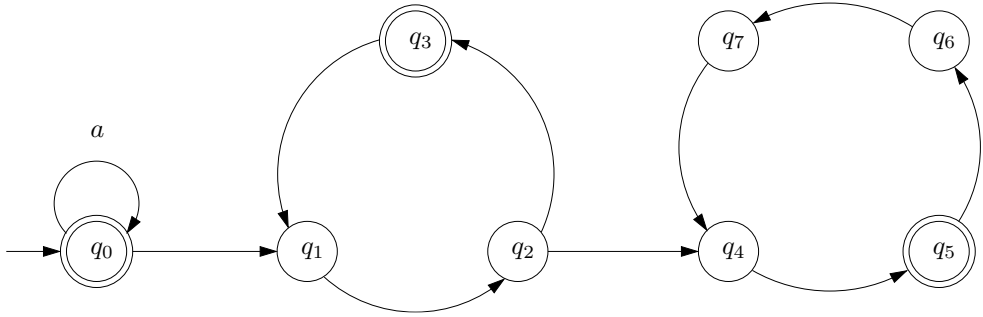 % 40%
    %\includegraphics{rl-automaton.pdf}
    % \begin{picture}(80,45)(0,0)
    %   \thinlines
    %   \node[Nmarks=ir](q0)(5,10){$q_0$}
    %   \node(q1)(25,10){$q_1$}
    %   \drawloop(q0){$a$}
    %   \drawedge(q0,q1){$b$}
    %   \node(q2)(45,10){$q_2$}
    %   \drawqbedge(q1,35,5,q2){$c$}
    %   \node[Nmarks=r](q3)(35,30){$q_3$}
    %   \drawqbedge(q2,55,25,q3){$c$}
    %   \drawqbedge(q3,15,25,q1){$b$}
    %   \node(q4)(65,10){$q_4$}
    %   \drawedge(q2,q4){$a$}
    % %
    %   \node[Nmarks=r](q5)(85,10){$q_5$}
    %   \drawqbedge(q4,75,5,q5){$a$}
    %   \node(q6)(85,30){$q_6$}
    %   \drawqbedge(q5,95,20,q6){$b$}
    %   \node(q7)(65,30){$q_7$}
    %   \drawqbedge(q6,75,43,q7){$c$}
    %   \drawqbedge(q7,55,20,q4){$a$}
    % \end{picture}
  \end{center}
  \caption{An rl-automaton}
  \label{fig:rla}
\end{figure}

One can convert any rlre to a DFA recognizing the same language.  For
example, Figure~\ref{fig:rla} displays a DFA for $a^* (bcc)^*
(bcaa)^*$ over $\Sigma = \{a, b, c\}$, where final states are circled.
The language $\Pfx (a^* (bcc)^* (bcaa)^*)$ is then recognized by the
same DFA, except that now all states are final.

This is general: $\Pfx (w_1^* w_2^* \ldots w_k^*)$ is always
recognizable by a DFA whose states are all final.  Let us therefore
call {\em rl-automaton\/} any such DFA.  Since all states are final,
we shall omit the $Fin$ component, and say that ${\mathcal A} =
(\Sigma, Q, \delta, q_0)$ itself is an rl-automaton.

Let us define the synchronized product.

\begin{defi}[Synchronized Product]
Let ${\mathfrak S}=(S,\stackrel{F}{\rightarrow}, \leq)$ be a
  complete functional transition system, and ${\mathcal A} = (F,
  %\times \Sigma,
  Q, \delta, q_0)$ be an rl-automaton on the same alphabet $F$.
  %\times \Sigma$.

  Define the {\em synchronized product\/} ${\mathfrak S} \times
  {\mathcal A}$ as the ordered functional transition system $(S \times
  Q, \stackrel{F'}{\rightarrow}, \leq')$, where $F'$ is the collection
  of all partial maps $f \bowtie \delta % a
  : (s, q) \mapsto (f (s),
  \delta (q,f
  %(f, a)
  ))$, for each $f \in F$
  %and $a \in \Sigma$
  such that $\delta (q, f
  %(f, a)
  )$ is defined for some $q \in Q$.  Let
  also $(s,q) \leq' (s',q')$ iff $s\leq s'$ and $q=q'$.

  Let $\pi_1$ be the morphism of transition systems defined as first
  projection on states; i.e., $\pi_1 (s,q) = s$ for all $(s,q) \in S \times Q$,
  $\pi_1 (f \bowtie \delta
  % a
  ) = f$ for all $f \in F$.
  %$(f,a) \in F \times \Sigma$.
\end{defi}

% is 
% {\em flat automaton\/} is a non-deterministic finite automaton
% ${\mathcal A} = (Q, \delta, q_0)$ that has no nested loop.  Formally,
% $Q$ is a finite set of so-called {\em control states\/}, $q_0 \in Q$
% is the initial state, and $\delta \subseteq Q \times F \times Q$ is
% the transition relation.  We write $q {\buildrel f \over\to} q'$ iff
% $(q, f, q') \in \delta$.  A {\em cycle\/} is any sequence of
% transitions $q_0 {\buildrel f_1\over\to} q_1 {\buildrel f_2\over\to}
% \ldots {\buildrel f_n\over\to} q_n = q_0$, where $n \geq 1$.  We also
% say that the set of states $\{q_1, \ldots, q_n\}$ is itself a cycle.
% Then flatness means that any two distinct cycles must be disjoint.  In
% a flat automaton, all states are assumed final, whence our slightly
% non-standard choice of not including a set of final states in the
% above definition of finite automata.  The language of any rlre is
% recognized by a flat automaton, even one of a very particular shape,
% illustrated on the left of Figure~\ref{fig:flattening}: a {\em linear
%   flat automaton\/} is a flat automaton ${\mathcal A} = (Q, \delta,
% q_0)$ where we can enumerate the states of $Q$ as $q_0, q_1, \ldots,
% q_n$, in such a way that whenever $(q_i, f, q_j)$ for some $f \in F$,
% then $i \leq j$.
\begin{lem}[Synchronized Product]
  \label{lem:strongly:flattable}
  Let ${\mathfrak S}=(S,\stackrel{F}{\rightarrow}, \leq)$ be a
  complete functional transition system, and ${\mathcal A} = (F,
  %\times \Sigma,
  Q, \delta, q_0)$ be an rl-automaton on the same alphabet $F$.
  %\times \Sigma$.

%  Define the {\em synchronized product\/} ${\mathfrak S} \times
%  {\mathcal A}$ as the ordered functional transition system $(S \times
%  Q, \stackrel{F'}{\rightarrow}, \leq')$, where $F'$ is the collection
%  of all partial maps $f \bowtie \delta % a
%  : (s, q) \mapsto (f (s),
%  \delta (q,f
%  %(f, a)
%  ))$, for each $f \in F$
%  %and $a \in \Sigma$
%  such that $\delta (q, f
%  %(f, a)
%  )$ is defined for some $q \in Q$.  Let
%  also $(s,q) \leq' (s',q')$ iff $s\leq s'$ and $q=q'$.
%
%  Let $\pi_1$ be the morphism of transition systems defined as first
%  projection on states; i.e., $\pi_1 (s,q) = s$ for all $(s,q) \in S \times Q$,
%  $\pi_1 (f \bowtie \delta
%  % a
%  ) = f$ for all $f \in F$.
%  %$(f,a) \in F \times \Sigma$.

  Then $({\mathfrak S} \times {\mathcal A}, \pi_1)$ is a continuous
  flattening of $\mathfrak S$.
\end{lem}
\proof First, the technical condition that $\delta (q, f
%(f, a)
)$ should be defined for some $q \in Q$ only excludes maps $f
\bowtie \delta
% a
$ with an empty domain, and is therefore benign.  This
technical condition is needed to define $\pi_1 (f \bowtie \delta
% a
)$ as $f$: formally, we define $\pi_1 (f')$ for any $f' \in F'$ by
letting $\pi_1 (f') (s)$ be the first component of the pair $f' (s,
q)$, where $q$ is some arbitrary state such that $\delta (q, f)$ is
defined, and let $\pi_1 (f') (s)$ be undefined otherwise; when $f' = f
\bowtie \delta
% a
$, such a $q$ exists by the technical condition, and this will yield
$f (s)$ when $s \in \dom f$, and will be undefined otherwise.  So
indeed $\pi_1 (f \bowtie \delta
% a
) = f$.

$(S \times Q, \leq')$ is easily seen to be a dcpo.  In fact, it is the
disjoint sum of finitely many copies of $S$, and as such, is a
continuous dcpo.  It is also well-ordered, as a finite disjoint sum of
well-ordered spaces.  So $S \times Q$ is a continuous dcwo.  Then we
check that $f \bowtie \delta
% a
$ is partial continuous.  Its domain is $\bigcup_{\substack{q \in Q\\
    \delta (q, f
%(f, a)
    )\text{ defined}}} \dom f \times \{q\}$, which is open.  Moreover
$f \bowtie \delta
% a
$ is clearly continuous for every $f \in F$:
% and $a \in \Sigma$:
for any directed family ${(s_i, q_i)}_{i \in I}$ in $\dom (f
\bowtie \delta
% a
)$, first all $q_i$s must be equal, say $q_i = q \in Q$, and second
${(s_i)}_{i \in I}$ must be directed in $\dom f$.  So $f (\lub \{s_i
\mid i \in I\}) = \lub \{f (s_i) \mid i \in I\}$, whence $(f
\bowtie \delta
% a
) (\lub \{(s_i, q) \mid i \in I\}) = (\lub \{f (s_i) \mid i \in I\},
\delta (q,f
%(f,a)
)) = \lub \{(f \bowtie \delta
% a
) (s_i, q) \mid i \in I\}$.  That $\pi_1$ is continuous is clear as
well.

Finally, the language of fireable transitions in ${\mathfrak S} \times
{\mathcal A}$ is contained in the language of $\mathcal A$, which is
of the form $\Pfx (w_1^* w_2^* \ldots w_k^*)$, hence bounded.  So
${\mathfrak S} \times {\mathcal A}$ is flat.  \qed

Strong flattenings are special: the decision to take the next action
$f \in F$ from state $(s, q)$ is dictated by the current control state
$q$ {\em only\/}, while ordinary flattenings allow more complex
decisions to be made.

We say that a transition system is strongly clover-flattable iff we
can require that the flat system ${\mathfrak S}_1$ is a synchronized
product, and the continuous morphism of transition systems $\varphi$
is first projection $\pi_1$:
\begin{defi}[Strongly Clover-Flattable]
  \label{defn:clover:flattable:strong}
  Let ${\mathfrak S}=(S,\stackrel{F}{\rightarrow})$ be a complete
  functional transition system.  We say that $({\mathfrak S},s_0)$ is
  \emph{strongly clover-flattable} iff there is an rl-automaton
  $\mathcal A$, say with initial state $q_0$, such that $cl
  (Cover_{\mathfrak S} (s_0)) = cl (\pi_1 \langle cl
  (Cover_{{\mathfrak S} \times {\mathcal A}} (s_0, q_0))\rangle)$.
\end{defi}

The following is then obvious.
\begin{lem}
  \label{lem:clover:flattable:easy}
  On complete functional transition systems, the implications
  ``strongly clover-flattable'' $\limp$ ``clover-flattable'' $\limp$
  ``weakly clover-flattable'' hold. \qed
\end{lem}
It is also easy to show that ``weakly clover-flattable'' also implies
``clover-flattable''.  However, we shall show something more general
in Theorem~\ref{thm:coverflat} below.

We show in Proposition~\ref{prop:coverflat:2} that {\bf
  Clover}$_{\mathfrak S} (s_0)$ can only terminate when $({\mathfrak
  S}, s_0)$ is strongly clover-flattable.  We shall require the
following lemma.  For notational simplicity, we equate words $g_1 g_2$
with compositions $g_2 \circ g_1$.
\begin{lem}
  \label{lemma:accel:approx}
  Let $\mathfrak S = (S, \stackrel F \rightarrow)$ be a complete
  functional transition system, and $s_0 \in F$.  Assume $\accel {g_1}
  \accel {g_2} \ldots \accel {g_n} (s_0)$ is defined, and in some open
  subset $U$ of $S$, for some $g_1, g_2, \ldots, g_n \in F$.  Then
  there are natural numbers $k_1, k_2, \ldots, k_n$ such that
  $g_1^{k_1} g_2^{k_2} \ldots g_n^{k_n} (s_0)$ is defined, and in $U$.
\end{lem}
\proof By induction on $n$.  This is clear if $n=0$.  Otherwise, let
$s = \accel {g_1} \accel {g_2} \ldots \accel {g_{n-1}} (s_0)$, so that
$\accel {g_n} (s)$ is defined and in $U$.  If $s < g_n (s)$, then
$\accel {g_n} (s) = \lub \{g_n^k (s) \mid k \in \nat\}$.  That the
latter is in the Scott-open $U$ implies that $g_n^{k_n} (s)$ is in $U$
for some $k_n \in \nat$.  If $s \not< g_n (s)$, then $\accel g_n (s) =
g_n (s)$, and we take $k_n=1$.  Let $V$ be the open
${(g_n^{k_n})}^{-1} (U)$.  (Note that, whereas $\accel {g_n}$ is not
partial continuous in general, $g_n^{k_n}$ is.)  So $s = \accel {g_1}
\accel {g_2} \ldots \accel {g_{n-1}} (s_0)$ is in $V$, in each case.
We apply the induction hypothesis and obtain the existence of $k_1,
k_2, \ldots, k_{n-1}$ such that $g_1^{k_1} g_2^{k_2} \ldots
g_{n-1}^{k_{n-1}} (s_0)$ is defined and in $V$.  Hence $g_1^{k_1}
g_2^{k_2} \ldots g_n^{k_n} (s_0)$ is defined, and in $U$, by
definition of $V$.  \qed

\begin{prop}
  \label{prop:coverflat:2}
  Let $\mathfrak S$ be an $\infty$-effective complete WSTS.  If {\bf
    Clover}$_{\mathfrak S}$ terminates on $s_0$, then $({\mathfrak
    S},s_0)$ is strongly clover-flattable.
\end{prop}
\proof Write $\mathfrak S$ as $(S, \stackrel{F}{\rightarrow}, \leq)$.
Assume that {\bf Clover}$_{\mathfrak S}$ terminates on $s_0$.  Then it
returns some finite set $A$ such that $A = Clover_{\mathfrak S} (s_0)$
by Theorem~\ref{cor}.  Enumerate the elements $a_1$, \ldots, $a_k$ of
$A$.  Each element $a_i$ of $A$, $1\leq i\leq k$, is obtained as
$\accel {g_{i1}} \accel {g_{i2}} \ldots \accel {g_{in_i}} (s_0)$,
where each $g_{ij}$ is in $F^*$.

Build a DFA for the language ${\mathcal L} = g_{1 1}^* g_{1 2}^*
\ldots g_{1 n_{1}}^* g_{2 1}^* g_{2 2}^* \ldots g_{2 n_{2}}^* \ldots
g_{k 1}^* g_{k 2}^* \ldots g_{k n_{k}}^*$.  Make all its states final,
so as to obtain an rl-automaton $\mathcal A$, with initial state
$q_0$.

We must show that $cl (Cover_{\mathfrak S} (s_0)) = cl (\pi_1 \langle
cl (Cover_{{\mathfrak S} \times {\mathcal A}} (s_0, q_0))\rangle)$,
i.e., that $\dc A = cl (\pi_1 \langle Cover_{{\mathfrak S} \times
  {\mathcal A}} (s_0, q_0))\rangle)$.

The inclusion from right to left is obvious: for any state $(s, q)$
that is reachable from $\dc (s_0, q_0)$ in ${\mathfrak S} \times
{\mathcal A}$, $s$ is reachable from $\dc s_0$ in $\mathfrak S$.  So
$\pi_1 \langle Post_{\mathfrak S}^* (\dc s_0) \rangle \subseteq
Post_{{\mathfrak S} \times {\mathcal A}}^* (\dc (s_0, q_0))$.  Taking
downward closures yields $\pi_1 \langle Cover_{\mathfrak S}^* (s_0)
\rangle \subseteq Cover_{{\mathfrak S} \times {\mathcal A}}^* (s_0,
q_0)$, and taking closures yields $cl (\pi_1 \langle Cover_{{\mathfrak
    S} \times {\mathcal A}} (s_0, q_0))\rangle) \subseteq cl
(Cover_{{\mathfrak S} \times {\mathcal A}}^* (s_0, q_0)) = \dc A$
(using Theorem~\ref{cor} and Proposition~\ref{prop:clover}).

The other inclusion reduces to showing that for every $i$, $1\leq
i\leq k$, the $i$th element $a_i$ of $A$ is in $cl (\pi_1 \langle
Cover_{{\mathfrak S} \times {\mathcal A}} (s_0, q_0))\rangle)$.  It is
equivalent to show that every open subset $U$ containing $a_i$
intersects $\pi_1 \langle Cover_{{\mathfrak S} \times {\mathcal A}}
(s_0, q_0))\rangle$.  By Lemma~\ref{lemma:accel:approx}, there are
natural numbers $k_1, k_2, \ldots, k_{n_i}$ such that $g_{i1}^{k_1}
g_{i2}^{k_2} \ldots g_{in}^{k_n} (s_0)$ is defined and in $U$.  Since
the word $g_{i1}^{k_1} g_{i2}^{k_2} \ldots g_{in}^{k_n}$ is in the
language $\mathcal L$, $g_{i1}^{k_1} g_{i2}^{k_2} \ldots g_{in}^{k_n}
(s_0)$ is the first component of some pair reachable from $(s_0, q_0)$
in ${\mathfrak S} \times {\mathcal A}$.  In particular, $g_{i1}^{k_1}
g_{i2}^{k_2} \ldots g_{in}^{k_n} (s_0)$ is in $\pi_1 \langle
Cover_{{\mathfrak S} \times {\mathcal A}} (s_0, q_0))\rangle$.  So $U$
intersects $\pi_1 \langle Cover_{{\mathfrak S} \times {\mathcal A}}
(s_0, q_0))\rangle$, as claimed.  \qed

We now loop the loop and show that {\bf Clover}$_{\mathfrak S}$
terminates on $s_0$ whenever $({\mathfrak S},s_0)$ is weakly
clover-flattable (Theorem~\ref{thm:coverflat} below).  This may seem
obvious.  In particular, if $({\mathfrak S},s_0)$ is clover-flattable,
then accelerate along the loops from ${\mathfrak S}_1$, where
${\mathfrak S}_1$, $\varphi$ is a continuous flattening of ${\mathfrak
  S}$.  The difficulty is that we {\em cannot\/} actually choose to
accelerate whenever we want: the {\bf Clover}$_{\mathfrak S}$
procedure decides by itself when it should accelerate, independently
of any flattening whatsoever.

There is an added difficulty, in the sense that one should also check
that lub-accelerations, as they are used in {\bf Clover}$_{\mathfrak
  S}$, are enough to reach all required least upper bounds.  The key
point is the following lemma, which asserts the existence of finitely
many subsequences $g^{p_j + \ell q_j} (s)$, $\ell \in \nat$, whose
exponents form infinite arithmetic progressions, and which generate
all possible limits of directed families of elements of the form $g^n
(s)$, $n \in \nat$, except possibly for finitely many isolated points.

This is the point in our study where progress is needed.  Indeed, we
require $S$ to be wpo to pick $k$ and $m$ in the proof below.
\begin{lem}
  \label{lemma:accel:partition}
  Let $S$ be a dcwo, $g : S \to S$ a partial monotonic map, and $s \in
  S$.  Consider the family $G$ of all elements of the form $g^n (s)$,
  for those $n \in \nat$ such that this is defined.  Then there are
  finitely many directed subfamilies $G_0$, $G_1$, \ldots, $G_{m-1}$
  of $G$ such that:
  \begin{enumerate}[\em(1)]
  \item $cl (G) = \bigcup_{j=0}^{m-1} cl (G_j) = \dc \{\lub (G_0), \lub
    (G_1), \ldots, \lub (G_{m-1})\}$;
%   \item $G_j \cap \dc \bigcup{\substack{k=1\\k\neq j}}^m cl (G_k)$ for
%     all $j$, $1\leq j\leq m$;
  \item each $G_j$ is either a one-element set $\{g^{p_j} (s)\}$,
    where $p_j \in \nat$, or is a chain of the form $\{g^{p_j + \ell
      q_j} (s) \mid \ell \in \nat\}$, where $p_j \in \nat$, $q_j \in
    \nat \setminus \{0\}$, and $g^{p_j} (s) < g^{p_j+q_j} (s)$;
  \item for every $j$, $0\leq j <m$, $s \not< g^{p_j} (s)$.
  \end{enumerate}
\end{lem}
\proof First, the claim is obvious if $G$ is finite, in which case we
just take $G_1$, \ldots, $G_m$ to consist of the sets $\{s_1\}$,
\ldots, $\{s_m\}$, where $G = \{s_1, \ldots, s_m\}$.  Write $s_j$ as
$g^{p_j} (s)$, and note that it cannot be the case that $s < g^{p_j}
(s)$, otherwise $g^{ip_j} (s)$ would be defined for all $i \in \nat$
(an easy induction on $i$, using the fact that the domain of $g^{p_j}$
is upward-closed), contradicting the fact that $G$ is finite.  So
condition~(3) holds.

So assume $G$ is infinite, i.e., $g^n (s)$ is defined for arbitrarily
large values of $n$.  Whenever $g^n (s)$ is defined, $g^m (s)$ is,
too, for all $m < n$.  So $g^n (s)$ is defined for all $n \in \nat$,
and $G = \{g^n (s) \mid n \in \nat\}$.  Since $S$ is wpo, for some $k,
m \in \nat$ with $k < m$, $g^k (s) \leq g^{m} (s)$.  We pick a minimal
$k$ such that $g^k (s) \leq g^m (s)$ for some $m > k$; and given $k$,
we pick a minimal $m > k$ such that $g^k (s) \leq g^m (s)$.

Let $G_0 = \{s\}$, $G_1 = \{g (s)\}$, \ldots, $G_{k-1} = \{g^{k-1}
(s)\}$, $G_k = \{g^{k + i (m-k)} (s) \mid i \in \nat\}$, $G_{k+1} =
\{g^{k+1+i (m-k)} (s) \mid i \in \nat\}$, \ldots, $G_{m-1} =
\{g^{m-1+i (m-k)} (s) \mid i \in \nat\}$.

Each $G_j$ is directed.  This is clear when $j < k$.  Otherwise, since
$g^k (s) \leq g^m (s)$ and $g$ is partial monotonic, we obtain
$g^{j+i(m-k)} (s) = g^{j-k + i(m-k)} (g^k (s)) \leq g^{j-k + i(m-k)}
(g^m (s)) = g^{j+ (i+1)(m-k)} (s)$.  So $G_j = {(g^{j + i (m-k)}
  (s))}_{i \in \nat}$ is an increasing chain.

Condition~(2) is satisfied: $G_j$ is a one-element set when $0\leq j <
k$, or when $k \leq j < m$ and $g^j (s) = g^{j+m-k} (s)$, i.e., when
the first two elements of $G_j$ are equal; indeed, in the latter case
$g^{j+i(m-k)} (s) = g^{i(m-k)} (g^j (s)) = g^{i(m-k)} (g^{j+m-k} (s))
= g^{j+(i+1)(m-k)} (s)$, so all elements of the sequence coincide.
Otherwise, i.e., if $k \leq j < m$ and $g^j (s) \neq g^{j+m-k} (s)$
(in which case $g^j (s) < g^{j+m-k} (s)$, since $g^{j+i(m-k)} (s) \leq
g^{j+(i+1)(m-k)} (s)$ for all $i$), let $p_j = j$ and $q_j = m-k$.

Let us establish condition~(1).  First, $G = \bigcup_{j=0}^{m-1} G_j$.
In particular, $G_j \subseteq G$, so $cl (G_j) \subseteq cl (G)$ for
all $j$, whence $\bigcup_{j=0}^{m-1} cl (G_j) \subseteq cl (G)$.
%We show the converse inclusion as follows.

Next, let $s_j = \lub (G_j)$ for all $j$, $0\leq j< m$.  This exists
because $G_j$ is a chain, hence is directed, and $S$ is a dcpo.  The
finite union $\bigcup_{j=0}^{m-1} \dc s_j$ is closed, and contains
$\bigcup_{j=0}^{m-1} G_j = G$, so it contains $cl (G)$.  Conversely,
the definition of $s_j$ makes it clear that $s_j \in cl (G_j)
\subseteq cl (G)$.  So $cl (G) = \bigcup_{j=0}^{m-1} \dc s_j = \dc
\{s_0, s_1, \ldots, s_{m-1}\}$.

% Fix a subset $J$ of $\{0, 1, \ldots, m-1\}$ so that $s_j$, $j \in J$,
% is an enumeration of the maximal elements among $s_0, s_1, \ldots,
% s_{m-1}$.  In particular, the elements $s_j$, $j \in J$ are pairwise
% incomparable, and $cl (G) = \dc \{s_j \mid j \in J\}$.

% We observe that, for every $j \in J$, the non-decreasing sequence $g^j
% (s)$, $g^{j+m-k} (s)$, \ldots, $g^{j+i(m-k)} (s)$, \ldots{} must
% eventually fall outside $\bigcup_{j' \in J, j' \neq j} \dc s_{j'}$.
% In other words, there is an index $i_j$ such that $g^{j+i(m-k)} (s)$
% is outside $\bigcup_{j' \in J, j' \neq j} \dc s_{j'}$ for all $i \geq
% i_j$.  Indeed, otherwise every $g^{j+i(m-k)} (s)$, $i\in \nat$, would
% be in the closed set $\bigcup_{j' \in J, j' \neq j} \dc s_{j'}$, hence
% also its least upper bound $s_j$; this would contradict the fact that
% the elements $s_j$, $j \in J$ are pairwise incomparable.

Take any element $x$ in $cl (G)$.  Since $x \in cl (G)$, $x \leq s_j$
for some $j$, $0\leq j < m$.  However, $s_j \in cl (G_j)$, and $cl
(G_j)$ is downward-closed, so $x \in \bigcup_{j=0}^{m-1} cl (G_j)$.
So $cl (G) \subseteq \bigcup_{j=0}^{m-1} cl (G_j)$.  So condition~(1)
holds.

Finally, assume condition~(3) failed.  Then $s < g^j (s)$ for some
$j$, $0\leq j< m$.  Certainly $j \neq 0$, since $g^0 (s)=s$.  By the
minimality of $k$ such that $g^k (s) \leq g^m (s)$ for some $m>k$,
$k=0$.  By the minimality of $m$, $m \leq j$.  But this contradicts
$j<m$.  \qed

\begin{prop}
  \label{prop:coverflat:1}
  Let $\mathfrak S$ be an $\infty$-effective complete WSTS.  Assume
  that $({\mathfrak S},s_0)$ is weakly clover-flattable.  Then {\bf
    Clover}$_{\mathfrak S}$ terminates on $s_0$.
\end{prop}
\proof Let ${\mathfrak S}_1$, $\varphi$ be a continuous flattening of
${\mathfrak S}$, and $s_1$ be a state of $\mathfrak S_1$ such that
$\varphi (s_1) \leq s_0$ and $Cover_{{\mathfrak S}}(s_0) \subseteq \dc
\varphi \langle Cover_{{\mathfrak S}_1}(s_1) \rangle$, i.e.,
$Clover_{{\mathfrak S}}(s_0) \leq^\flat \varphi \langle
Clover_{{\mathfrak S_1}}(s_1)\rangle$.  Write $\mathfrak S_1$ as
$(S_1, \stackrel{F_1}{\rightarrow}, \leq)$.  Since $\mathfrak S_1$ is
flat, every $g_1 \in F_1$ is in $w_1^* w_2^* \ldots w_m^*$, for some
fixed sequence $w_1, w_2, \ldots, w_m \in F_1^*$.

Extend the action of $\varphi : F_1 \to F$ on words by $\varphi (f_1
f_2 \ldots f_p) = \varphi (f_1) \varphi (f_2) \ldots \varphi (f_p)$.
Thus $\varphi (w_1)$, \ldots, $\varphi (w_m)$ are defined.

Consider first $\varphi (w_1)$.  Apply
Lemma~\ref{lemma:accel:partition} with $g = \varphi (w_1)$ and
$s=s_0$, and get finitely many subfamilies of $G_0$, $G_1$, \ldots,
$G_{m-1}$ of $G = \{\varphi (w_1)^n (s_0) \mid n \in \nat, \varphi
(w_1)^n (s_0) \text{ is defined}\}$ satisfying the conditions given in
the Lemma.

For each $j$ such that $G_j$ is a one-element set, say $G_j =
\{\varphi (w_1)^n (s_0)\}$, observe that {\bf Clover}$_{\mathfrak S}$
will eventually select the pair $(\varphi (w_1)^n, s_0)$ at line 2.(a)
by fairness, and add $\accel {(\varphi (w_1)^n)} (s_0)$ to $A$.  By
condition~(3), $s_0 \not< \varphi (w_1)^n (s_0)$, so $\accel {(\varphi
  (w_1)^n)} (s_0) = \varphi (w_1)^n (s_0)$.  So {\bf
  Clover}$_{\mathfrak S}$ will eventually add $\varphi (w_1)^n (s_0) =
\lub (G_j)$ to $A$.

Still taking the notations of the Lemma, for every $j$ such that $G_j$
contains more than one element, {\bf Clover}$_{\mathfrak S}$ will
eventually select the pair $(\varphi (w_1)^{p_j}, s_0)$, adding
$\accel {(\varphi (w_1)^{p_j})} (s_0)$ to $A$.  Using condition~(3) as
above, one sees that $\accel {(\varphi (w_1)^{p_j})} (s_0) = \varphi
(w_1)^{p_j} (s_0)$.  Then, by fairness again (and this is the
important point in the proof, where lub-acceleration is needed), {\bf
  Clover}$_{\mathfrak S}$ will eventually select the pair $(\varphi
(w_1)^{q_j}, \varphi (w_1)^{p_j} (s_0))$, and therefore add $\accel
{(\varphi (w_1)^{q_j})} (\varphi (w_1)^{p_j} (s_0))$ to $A$.  By
condition~(2), $\accel {(\varphi (w_1)^{q_j})} (\varphi (w_1)^{p_j}
(s_0))$ is just $\lub \{\varphi (w_1)^{p_j+\ell q_j} (s_0) \mid \ell
\in \nat\} = \lub (G_j)$.

Let again $A_n$ be the value of the set $A$, computed by the procedure
{\bf Clover}$_{\mathfrak S}$ on input $s_0$, after $n$ iterations of
the while statement at line $2$.  Let $A = \bigcup_{n \in \nat} A_n$.
We have just shown that at some step, say $n_1$, {\bf
  Clover}$_{\mathfrak S}$ will have added enough elements to $A$ so
that every element of the form $\varphi (w_1)^{k_1} (s_0)$, $k_1 \in
\nat$ (provided this is defined), is below some element of $A_{n_1}$.

Let us proceed with $\varphi (w_2)$.  Fix an arbitrary element $s$ of
$A_{n_1}$, and apply Lemma~\ref{lemma:accel:partition} with $g =
\varphi (w_2)$.  Proceeding as above, we observe that there is an $n_2
\geq n_1$ such that every element of the form $\varphi (w_2)^{k_2}
(s)$, $n \in \nat$, is below some element of $A_{n_2}$.  Since $s$ is
arbitrary in $A_{n_1}$, we conclude that every element of the form
$\varphi (w_2)^{k_2} (\varphi (w_1)^{k_1} (s_0))$, $k_1, k_2 \in
\nat$, is below some element of $A_{n_2}$.

We now induct on $i$, $1\leq i\leq m$, to show similarly that there is
an $n_i \in \nat$ such that every element of the form $\varphi
(w_i)^{k_i} (\varphi (w_{i-1})^{k_{i-1}} (\ldots \varphi (w_1)^{k_1}
(s_0)))$, where $k_1, \ldots, k_i \in \nat$, is below some element of
$A_{n_i}$.

In particular, for $i=m$, writing $n$ for $n_m$: $(*)$ there is an $n
\in \nat$ such that every element of the form $\varphi (w_m)^{k_m}
(\varphi (w_{m-1})^{k_{m-1}} (\ldots \varphi (w_1)^{k_1} (s_0)))$,
where $k_1, \ldots, k_m \in \nat$, is below some element of $A_n$.  We
claim that {\bf Clover}$_{\mathfrak S} (s_0)$ must stop after step
$n$.

Let $U$ be the (open) complement of the closed set $\dc A_n$, and
assume that $U$ intersects $\dc Clover_{{\mathfrak S} }(s_0)$.  Then
$U$ must also intersect $\dc \varphi \langle Clover_{{\mathfrak
    S_1}}(s_1) \rangle$, hence $\varphi \langle Clover_{{\mathfrak
    S_1}}(s_1) \rangle$.  (Remember that open subsets are
upward-closed.)  So $\varphi^{-1} (U)$ intersects $Clover_{{\mathfrak
    S_1}}(s_1)$, whence $\varphi^{-1} (U)$ intersects $\dc
Clover_{{\mathfrak S_1}}(s_1)$, since $\varphi^{-1} (U)$ is
upward-closed, using the fact that $U$ is and that $\varphi$ is
monotonic. By Proposition~\ref{prop:clover}, $\varphi^{-1} (U)$
intersects $cl (Cover_{{\mathfrak S_1}}(s_1))$. Since $\varphi$ is
continuous, $\varphi^{-1} (U)$ is open. We now use the fact that an
open intersects the closure of a set iff it intersects that set. So
$\varphi^{-1} (U)$ must intersect $Cover_{{\mathfrak S_1}}(s_1)$. So
$U$ intersects $\varphi \langle Cover_{{\mathfrak S}_1} (s_1)\rangle$,
say at $a$. In particular, there is an $a_1 \in S_1$ such that $a \leq
\varphi (a_1)$, and $a_1 \leq w_1^{k_1} w_2^{k_2} \ldots w_m^{k_m}
(s_1)$, for some natural numbers $k_1$, $k_2$, \ldots, $k_m$.

Since $a \leq \varphi (a_1) \leq \varphi (w_1^{k_1} w_2^{k_2} \ldots
w_m^{k_m}) (\varphi (s_1)) \leq \varphi (w_1^{k_1} w_2^{k_2} \ldots
w_m^{k_m}) (s_0) = \varphi (w_m)^{k_m} \allowbreak (\varphi
(w_{m-1})^{k_{m-1}} \allowbreak (\ldots \varphi (w_1)^{k_1} (s_0)))$,
$a$ is in $\dc A_n$ by $(*)$.  But this contradicts the fact that $a
\in U$.  So the complement $U$ of $\dc A_n$ does not intersect $\dc
Clover_{{\mathfrak S} }(s_0)$, i.e., $\dc Clover_{\mathfrak S} (s_0)
\subseteq \dc A_n$.

By Proposition~\ref{prop:An:clover}, the converse inclusion holds.  We
conclude that the procedure {\bf Clover}$_{\mathfrak S}$ stops after
the $n$th turn of the loop, because of the fixpoint test at line 2.
\qed

Putting together Lemma~\ref{lem:clover:flattable:easy},
Proposition~\ref{prop:coverflat:2}, and
Proposition~\ref{prop:coverflat:1}, we obtain:
\begin{thm}[Main Theorem]
  \label{thm:coverflat}
  Let ${\mathfrak S}$ be an $\infty$-effective complete WSTS. The
  following statements are equivalent:
  \begin{enumerate}[\em(1)]
  \item $({\mathfrak S}, s_0)$ is clover-flattable;
  \item $({\mathfrak S}, s_0)$ is weakly clover-flattable;
  \item $({\mathfrak S}, s_0)$ is strongly clover-flattable;
  \item {\bf Clover}$_{\mathfrak S} (s_0)$ terminates.  \qed
  \end{enumerate}
\end{thm}

%!!! penser a montrer que Clover est aussi correcte meme pour les executions infinies... en fait peut-etre pas (depend du fait que A = bigcup des A_n est fini)

\subsection{Cover-flattability (without the ``l'' in ``Cover'')}

Turning to non-complete WSTS, we define:
\begin{defi}[Monotonic Flattening]
  \label{defn:flatten:mono}
  Let ${\mathfrak X}_2=(X_2,\stackrel{F_2}{\rightarrow}, \leq_2)$ be
  an ordered functional transition system.  A flattening $({\mathfrak
    X_1}, \varphi)$ of ${\mathfrak X}_2$ is {\em monotonic\/} iff:
  \begin{enumerate}[(1)]
  \item ${\mathfrak X}_1= (X_1, \stackrel{F_1}{\rightarrow}, \leq_1)$
    is an ordered functional transition system;
  \item and $\varphi : X_1 \to X_2$ is {\em monotonic\/}.
  \end{enumerate}
\end{defi}

\begin{defi}[Cover-Flattable]
  \label{defn:cover:flattable}
  Let $\mathfrak X$ be an ordered functional transition system, and
  $x_0$ be a state.  We say that $({\mathfrak X},x_0)$ is
  \emph{cover-flattable} iff there is a monotonic flattening
  $({\mathfrak X}_1, \varphi)$ of ${\mathfrak X}$, and a state $x_1$
  of $\mathfrak X_1$ such that:
  \begin{enumerate}[(1)]
  \item $\varphi (x_1) = x_0$;
  \item $Cover_{\mathfrak X} (x_0) = \dc \varphi \langle
    Cover_{{\mathfrak X}_1} (x_1)\rangle$.
  \end{enumerate}
\end{defi}

\begin{defi}[Weakly Cover-Flattable]
  \label{defn:cover:flattable:weak}
  Let $\mathfrak X$ be an ordered functional transition system, and
  $x_0$ be a state.  We say that $({\mathfrak X},x_0)$ is \emph{weakly
    cover-flattable} iff there is a monotonic flattening $({\mathfrak
    X}_1, \varphi)$ of ${\mathfrak X}$, and a state $x_1$ of
  $\mathfrak X_1$ such that:
  \begin{enumerate}[(1)]
  \item $\varphi (x_1) \leq x_0$;
  \item and $Cover_{\mathfrak X} (x_0) \subseteq \dc \varphi \langle
    Cover_{{\mathfrak X}_1} (x_1)\rangle$.
  \end{enumerate}
\end{defi}

\begin{defi}[Strongly Cover-Flattable]
  \label{defn:cover:flattable:strong}
  Let ${\mathfrak X}=(X,\stackrel{F}{\rightarrow})$ be an ordered
  functional transition system.  We say that $({\mathfrak X},x_0)$ is
  \emph{strongly cover-flattable} iff there is an rl-automaton
  $\mathcal A$, say with initial state $q_0$, such that
  $Cover_{\mathfrak X} (x_0) = \pi_1 \langle Cover_{{\mathfrak X}
    \times {\mathcal A}} (x_0, q_0)\rangle$.
\end{defi}

\begin{thm}
  \label{thm:cover:flat}
  Let ${\mathfrak X} = (X, \stackrel F \rightarrow, \leq)$ be an
  $\omega^2$-WSTS that is $\infty$-effective, in the sense that
  $\widehat {\mathfrak X}$ is $\infty$-effective, i.e., that $\accel
  {(\Sober g)}$ is computable for every $g \in F^*$.  The following
  statements are equivalent:
  \begin{enumerate}[\em(1)]
  \item $({\mathfrak X}, x_0)$ is cover-flattable;
  \item $({\mathfrak X}, x_0)$ is weakly cover-flattable;
  \item $({\mathfrak X}, x_0)$ is strongly cover-flattable;
  \item $(\widehat {\mathfrak X}, \eta_X (x_0))$ is (weakly, strongly)
    clover-flattable;
  \item {\bf Clover}$_{\widehat {\mathfrak X}} (\eta_X (x_0))$
    terminates.
  \end{enumerate}
  In this case, {\bf Clover}$_{\widehat {\mathfrak X}} (\eta_X (x_0))$
  returns the clover $A = Clover_{\mathfrak S} (s_0)$, and this is a
  finite description of the cover, in the sense that $Cover_{\mathfrak
    X} (x_0) = \eta_X^{-1} (\dc A)$.
\end{thm}
\proof First, that {\bf Clover}$_{\widehat {\mathfrak X}} (\eta_X
(x_0))$ computes the clover $A$ is Theorem~\ref{cor}, and the fact
that $Cover_{\mathfrak X} (x_0) = \eta_X^{-1} (\dc A)$, by
Proposition~\ref{prop:completion:clover}.  If we equate $X$ with
$\eta_X \langle X \rangle$, the latter means that the cover is just $X
\cap \dc A$.

Next, (4) is equivalent to (5), by Theorem~\ref{thm:coverflat}.
Note in particular that $\widehat {\mathfrak X}$ is a complete WSTS
by Theorem~\ref{thm:omega2}, and is $\infty$-effective by assumption.

The implications (1) $\limp$ (2) and (3) $\limp$ (1) are clear.  For
the latter, note that, since $Cover_{{\mathfrak X} \times {\mathcal
    A}} (x_0, q_0)$ is downward-closed, $\pi_1 \langle
Cover_{{\mathfrak X} \times {\mathcal A}} (x_0, q_0)\rangle = \dc
\pi_1 \langle Cover_{{\mathfrak X} \times {\mathcal A}} (x_0,
q_0)\rangle$, and take $\varphi = \pi_1$.

We now show that (2) implies (4), i.e., that if $({\mathfrak X}, x_0)$
is weakly cover-flattable, then $(\widehat {\mathfrak X}, \eta_X
(x_0))$ is weakly clover-flattable.  So let ${\mathfrak X}_1 = (X_1,
\stackrel {F_1} \rightarrow, \leq)$, $\varphi$ and $x_1$ as in
Definition~\ref{defn:cover:flattable:weak}.  In particular, $\varphi
(x_1) \leq x_0$ and $Cover_{\mathfrak X} (x_0) \subseteq \dc \varphi
\langle Cover_{{\mathfrak X}_1} (x_1)\rangle$.  Let $S_1$ be the ideal
completion $\Idl ({\mathfrak X}_1)$, with inclusion as ordering, and
define the complete transition system ${\mathfrak S}_1 = (S_1,
\stackrel {F'_1} \rightarrow, \subseteq)$, where $F'_1 = \{\Idl (f)
\mid f \in F_1\}$.  $\Idl (f)$ is the partial continuous function that
maps every ideal $D$ such that $D \cap \dom f \neq \emptyset$ to $\dc
f \langle D\rangle$.
% (We check that $\Idl (f)$ is partial continuous: its domain is
% clearly Scott-open, $\Idl (f)$ is monotonic, and for every
% directed family ${(D_i)}_{i \in I}$ of ideals,
% $\Idl (f) (\bigcup_{i \in I} D_i) = \dc \bigcup_{i \in I} f \langle D_i \rangle
% = \bigcup_{i \in I} \dc f \langle D_i \rangle$.)
Remember that $\widehat {\mathfrak X} = \Idl ({\mathfrak X})$.  Define
$\varphi' : S_1 \to \widehat{\mathfrak X}$ as $\Idl (\varphi)$: this
is continuous.  On transitions, $\varphi'$ maps $\Idl (f)$ to $\Idl
(\varphi (f))$: this is well-defined, as one can recover $f$ from
$\Idl (f)$, by the fact that $f (x) = \lub (\Idl (f) (\dc x))$.  So
$({\mathfrak S}_1, \varphi')$ is a continuous flattening of $\widehat
{\mathfrak X}$.  Let $s_1 = \dc x_1$, $s_0 = \dc x_0$.  We claim that
$\varphi' (s_1) \subseteq s_0$, and that $Cover_{\widehat {\mathfrak
    X}} (s_0) \subseteq cl (\varphi' \langle cl (Cover_{{\mathfrak
    S}_1} (s_1))\rangle)$.  The first inequality is because $\varphi'
(s_1) = \Idl (\varphi) (\dc x_1) = \dc \varphi \langle \dc x_1 \rangle
= \dc \varphi (x_1) \subseteq \dc x_0 = s_0$, since $\varphi (x_1)
\leq x_0$.  For the second inequality, let $s$ be any element of
$Cover_{\widehat {\mathfrak X}} (s_0)$.  So $s \subseteq g (s_0)$ for
some $g \in {F'_1}^*$.  We observe that $\Idl$ is a functor, i.e.,
that $\Idl$ of the identity map is the identity, and that $\Idl
(g_1g_2) = \Idl (g_1) \Idl (g_2)$ for all $g_1$, $g_2$.  So, writing
$g$ as a composition $g_1 g_2 \ldots g_k$ of elements $g_i = \Idl
(h_i)$ of $F'_1$, $g$ equals $\Idl (h)$, where $h = h_1 h_2 \ldots h_k
\in F_1^*$.  It follows that $s \subseteq \Idl (h) (\dc x_0) = \dc h
(x_0)$.  Observe that $h (x_0) \in Cover_{\mathfrak X} (x_0) \subseteq
\dc \varphi \langle Cover_{{\mathfrak X}_1} (x_1)\rangle$, so $s
\subseteq \dc \varphi \langle Cover_{{\mathfrak X}_1} (x_1)\rangle$.
In particular, every element $x$ of the ideal $s$ is below some
element of the form $\varphi (f (x_1))$, $f \in F_1^*$.  We observe
that $x \in \varphi' (\Idl (f) (s_1))$: indeed, $\varphi' (\Idl (f)
(s_1)) = \Idl (\varphi) (\Idl (f) (s_1)) = \Idl (\varphi \circ f)
(s_1) = \dc (\varphi \circ f) \langle \dc x_1 \rangle = \dc \varphi (f
(x_1))$, and $x$ is in the latter since $x \leq \varphi (f (x_1))$.
>From $x \in \varphi' (\Idl (f) (s_1))$, and since $\Idl (f) (s_1) \in
Cover_{{\mathfrak S}_1} (s_1)$, we deduce that $x \in \varphi' \langle
Cover_{{\mathfrak S}_1} (s_1) \rangle$.  Since $x$ is arbitrary in
$s$, $s \subseteq \varphi' \langle Cover_{{\mathfrak S}_1} (s_1)
\rangle$, i.e., $s \in \dc \varphi' \langle Cover_{{\mathfrak S}_1}
(s_1) \rangle \subseteq cl (\varphi' \langle cl (Cover_{{\mathfrak
    S}_1} (s_1))\rangle)$.

Finally, we show that (4) implies (3), i.e., that if $(\widehat
{\mathfrak X}, \eta_X (x_0))$ is strongly clover-flattable, then
$({\mathfrak X}, x_0)$ is strongly cover-flattable.  Let $\mathcal A$
be an rl-automaton, with initial state $q_0$, such that $cl
(Cover_{\widehat {\mathfrak X}} (\eta_X (x_0))) = cl (\pi_1 \langle cl
(Cover_{{\widehat {\mathfrak X}} \times {\mathcal A}} (\eta_X (x_0),
q_0))\rangle)$.  We claim that $Cover_{\mathfrak X} (x_0) = \pi_1
\langle Cover_{{\mathfrak X} \times {\mathcal A}'} (x_0, q_0)\rangle$,
where $\mathcal A'$ is the automaton obtained from $\mathcal A$ by
replacing each $\Sober g$ transition by a $g$ transition, $g \in F$.
(Note by the way that the definition of $\Sober g$ is the same as that
of $\Idl (g)$ above.)  The inclusion from right to left is obvious, so
let us show that $Cover_{\mathfrak X} (x_0) \subseteq \pi_1 \langle
Cover_{{\mathfrak X} \times {\mathcal A}'} (x_0, q_0)\rangle$.  Let
$x$ be any element of $Cover_{\mathfrak X} (x_0)$.  So $x \leq g
(x_0)$ for some $g \in F$.  Then $x \in \dc g (x_0) = \dc g \langle
\dc x_0 \rangle = \Idl (g) (\eta_X (x_0))$, so $\dc x \in
Cover_{\widehat {\mathfrak X}} (\eta_X (x_0))$.  By assumption $\dc x$
is in $cl (\pi_1 \langle cl (Cover_{{\widehat {\mathfrak X}} \times
  {\mathcal A}} (\eta_X (x_0), q_0))\rangle)$.  We may simplify this
by observing that $cl (f \langle cl (A) \rangle) = cl (f \langle A
\rangle)$ for any continuous map $f$ and any subset $A$,
% (the hard direction $f \langle cl (A) \rangle \subseteq cl (f \langle A \rangle)$ is shown by letting $U$ be the open complement of $cl (f \langle A \rangle)$, and realizing that $U$ is contained in the complement of $f \langle cl (A) \rangle$: any element of $U$ of the form $f (x)$ is such
% that $x$ is in the open $f^{-1} (U)$, and the latter
% is the complement of $f^{-1} (cl (f \langle A\rangle))$, so $x$ is
% not in $A$)
so that $\dc x \in cl (\pi_1 \langle Cover_{{\widehat {\mathfrak X}}
  \times {\mathcal A}} (\eta_X (x_0), q_0)\rangle)$.  In $\widehat X =
\Idl (X)$, the closure $cl (A)$ of any downward-closed subset $A$ of
$\Idl (X)$ equals $\Lub (A)$, since $\Idl (X)$ is continuous.  It
follows that, if $\dc x \in cl (A)$, then $\dc x$ is the union of a
directed family ${(s_i)}_{i \in I}$ of elements of $A$; in particular,
$x$ is in some $s_i$, $i\in I$, i.e., $x$ is in some element (an
ideal) of $A$.  Taking $A = \pi_1 \langle Cover_{{\widehat {\mathfrak
      X}} \times {\mathcal A}} (\eta_X (x_0), q_0)\rangle$, $x$ is in
some ideal $s$ such that $(s, q) \in Cover_{{\widehat {\mathfrak X}}
  \times {\mathcal A}} (\eta_X (x_0), q_0)$ for some state $q$ of
$\mathcal A$.  That is, $s \subseteq \Sober g (\eta_X (x_0))$ for some
$g = g_1 g_2 \ldots g_k$, where $g_1, g_2, \ldots, g_k \in F$, and $q$
is the state obtained by reading the word $\Sober {g_1} \Sober {g_2}
\ldots \Sober {g_k}$ in $\mathcal A$ from $q_0$.  In particular, $q$
is also the state obtained by reading the word $g_1 g_2 \ldots g_k$ in
${\mathcal A}'$ from $q_0$.  And $s \subseteq \Sober g (\eta_X (x_0))$
means that $s \subseteq \dc g \langle \dc x_0 \rangle = \dc g (x_0)$,
so $x \in s$ implies $x \leq g (x_0)$.  In particular, $(x,q) \in
Cover_{{\mathfrak X} \times {\mathcal A}'} (x_0, q_0)$, so $x \in
\pi_1 \langle Cover_{{\mathfrak X} \times {\mathcal A}'} (x_0, q_0)
\rangle$.  \qed

By a slight abuse of language, say that a functional WSTS $\mathfrak S
= (S, \stackrel F \rightarrow, \leq)$ is cover-flattable iff
$(\mathfrak S, s_0)$ is cover-flattable for every initial state $s_0
\in S$.
\begin{cor}
  \label{cor:Petri:flat}
  Every Petri net, and every VASS, is cover-flattable.
\end{cor}
\proof The state space of a Petri net on $k$ places is $\nat^k$, that
of a VASS \cite{HP:VASS} is $Q \times \nat^k$, where $Q$ is a finite
set of control states.  We deal with the latter, as they are more
general.  Transitions of the VASS $\mathfrak X$ are of the form $f (q,
\vec x) = (q', \vec x + \vec b - \vec a)$, provided $\vec x \geq \vec
a$, and where $\vec a$, $\vec b$ are fixed tuples in $\nat^k$.  It is
easy to see that $\Sober f$ is defined by: $\Sober f (q, \vec x) =
(q', \vec x + \vec b - \vec a)$, provided $\vec x \geq \vec a$, this
time for all $\vec x \in \nat_\omega^k$.  So the completion $\widehat
{\mathfrak S}$ of the VASS is $\infty$-effective.  On these, the
Karp-Miller algorithm terminates \cite{KM:petri}, hence also the
generalized Karp-Miller algorithm of Section~\ref{sec:omega2:motiv}.
By Proposition~\ref{moreKM}, {\bf Clover}$_{\widehat {\mathfrak S}}$
terminates on any input $s_0 \in Q \times \nat_\omega^k$.  So
$\mathfrak X$ is cover-flattable, by Theorem~\ref{thm:cover:flat}.
\qed

\begin{cor}
  \label{cor:reset:notflat}
  There are reset Petri nets, and functional-lossy channel systems
  that are not cover-flattable.
\end{cor}
\proof One can again show that their completions are
$\infty$-effective, see Section~\ref{sec:omega2:eff}.  However the
cover is undecidable both for reset Petri nets and (functional-)lossy
channel systems $\mathfrak X$, so {\bf Clover}$_{\widehat {\mathfrak
    X}} (\eta_X (x_0))$ must fail to terminate for some initial state
$x_0$.  We conclude by Theorem~\ref{thm:cover:flat}.  \qed

\section{Application: Well Structured Counter Systems}
\label{sec:counter}

%complexité de la lub-accélération pour les B-VASS, data nets, ???
%%%%%%%%%%%%%%%%%%%%%%%%%%%%%%%%%%%%%%%%%%%%%%%%%%%%%%%%%
%\subsection{Completely Well-Structured Counter Systems}
%%%%%%%%%%%%%%%%%%%%%%%%%%%%%%%%%%%%%%%%%%%%%%%%%%%%%%%%%

We now demonstrate
% [jgl] supprime ``that'' (referee 2)
how the fairly large class of counter systems fits with our theory. We
show that counter systems composed of affine monotonic functions with
upward-closed definition domains are complete (strongly monotonic)
WSTS. This result is obtained by showing that every monotonic affine
function $f$ is continuous and its lub-acceleration $\accel f$ is
computable \cite{CFS-atpn2011}.  Moreover, we prove that it is
possible to decide whether a general counter system (given by a finite
set of Presburger relations) is a monotonic affine counter system, but
that one cannot decide whether it is a WSTS.

%%%%%%%%%%%%%%%%%%%%%%%%%%%%%%%%%%%%%%%%%%%%%%%%%%%%%%%%%%
%\subsubsection{Affine Counter Systems}
%%%%%%%%%%%%%%%%%%%%%%%%%%%%%%%%%%%%%%%%%%%%%%%%%%%%%%%%%%

\begin{defi}
  A \emph{relational counter system} (with $n$ counters), for short an
  \emph{$R$-counter system}, $\mathcal{C}$ is a tuple $\mathcal{C}=(Q,
  R,\rightarrow)$ where $Q$ is a finite set of control states,
  $R=\{r_1,r_2,...r_k\}$ is a finite set of Presburger relations $r_i
  \subseteq \mathbb{N}^n \times \mathbb{N}^n$ and $\rightarrow
  \subseteq Q \times R \times Q$.
% such that for every $q,q' \in \mathbb{Q}$, there
%exists at most one $r \in R$ such that $(q,r,q') \in \rightarrow$.
\end{defi}
We will consider a special case of Presburger relations, those which
allow us to encode the graph of affine functions. A (partial) function
$f: \mathbb{N}^n \longrightarrow \mathbb{N}^n$ is \emph{non-negative
  affine}, for short \emph{affine} if there exist a matrix $A \in
\mathbb{N}^{n \times n}$ with {\em non-negative coefficients\/} and a
vector $b \in \mathbb{Z}^n$ such that for all $\vec x \in \dom f,
f(\vec x)=A\vec x+\vec b$.
%   Note
% that $A$ has coefficients in $\mathbb{N}$, not in $\mathbb{Z}$,
% $\mathbb{Q}$ or $\mathbb{R}$.
When necessary, we will extend affine maps $f: \mathbb{N}^n
\longrightarrow \mathbb{N}^n$ by continuity to $f:
\mathbb{N}_{\omega}^n \longrightarrow \mathbb{N}_{\omega}^n$, by
$f(\lub_{i \in \nat} (\vec x_i))=\lub_{i \in \nat} (f(\vec x_i))$ for
every countable chain $(\vec x_i)_{i \in \nat}$ in $\nat^n$.
That is, we just write $f$ instead of $\Sober f$.

\begin{defi}
  \label{defn:ACS}
  An \emph{affine counter system} (with $n$ counters), a.k.a.\ an {\em
    ACS\/} $\mathcal{C}=(Q, R,\rightarrow)$ is a $R$-counter system
  where all relations $r_i$ are (partial) affine functions.
\end{defi}

The domain of maps $f$ in an affine counter system $ACS$ are
Presburger-definable. A reset/transfer Petri net is an $ACS$ where
every line or column of every matrix contains at most one non-zero
coefficient equal to $1$, and, all domains are upward-closed sets.  A
\emph{Petri net} is an ACS where all affine maps are translations with
upward-closed domains.

\begin{thm}
One can decide whether an effective relational counter system is an $ACS$.
\end{thm}
\proof The formula expressing that a relation is a function is a
Presburger formula, hence one can decide whether $R$ is the graph of a
function.  One can also decide whether the graph $G_f$ of a function
$f$ is monotonic because monotonicity of a Presburger-definable
function can be expressed as a Presburger formula. Finally, one can
also decide whether a Presburger formula represents an affine function
$f(\vec x)=A\vec x+\vec b$ with $A \in \mathbb{N}^{n \times n}$ and
$\vec b \in \mathbb{Z}^n$, using results by Demri {\em et al.\/}
\cite{DemriFGD06}.  \qed

%On the other side, it seems that the continuity and the dcwo properties
%are of the second order and they are probably undecidable in the Presburger
%logics. Hence we will use the usual ordering $\leq$ on $\mathbb{N}^n$ which
%is a continuous dcwo.\\

%Let us recall that a quasi ordered transition system
%${\mathfrak S}=(S,\rightarrow,\leq)$ is \emph{monotone} if for every $s,s',s_1
%\in S$ such that $s \rightarrow s'$ and $s_1 \geq s$, there exists an
%$s_1' \in S$ such that $s_1 \stackrel{*}{\rightarrow} s_1'$ and  $s_1'
%\geq s'$. We remark that the length of the path $s_1
%\stackrel{*}{\rightarrow} s_1'$ is not known in advance, and then 
For counter systems (which include Minsky machines), monotonicity is
undecidable.  Clearly, a counter system ${\mathfrak S}$ is
well-structured iff ${\mathfrak S}$ is monotonic: so there is no
algorithm to decide whether a relational counter system is a WSTS.
%   So
% define $k$-monotonicity, $k \geq 1$: ${\mathfrak S}$ is
% \emph{$k$-monotone} if for every $s,s',s_1 \in S$ such that $s
% \rightarrow s'$ and $s_1 \geq s$, there exists an $s_1' \in S$ such
% that $s_1 \stackrel{\leq k}{\rightarrow} s_1'$ and $s_1' \geq s'$;
% $k$-monotonicity can be expressed as a Presburger formula, and is
% therefore decidable for ACS.
However, an ACS is strongly monotonic iff each map $f$ is partial
monotonic; this is equivalent to requiring that $\dom f$ is
upward-closed, since all matrices $A$ have non-negative coefficients.
This is easily cast as Presburger formula, and therefore decidable.
%%%%

\begin{prop}
  There is an algorithm to decide whether an $ACS$ is a strongly
  monotonic WSTS.
\end{prop}
\proof The strong monotony of an ACS $\mathcal{C}$ means that every
function of $\mathcal{C}$ is monotonic and this can be expressed by a
Presburger formula saying that all the (Presburger-definable)
definition domains are upward-closed (the matrices are known to be
positive).  \qed

%\subsection{Characterization of Completely Well-Structured Counter Systems}

% a verifier%%%%%%%%%
%Let us note that for every Presburger function
%$f_i$, the domain of $f_i$  is Presburger definable;
%moreover, when $f_i$ is monotone, its domain is an upward-closed set.
%%
%
%We now prove that it is possible to compute $\accel f(x)=\lub(f^p(x))_{p \geq
%  0}$ for any monotone affine function. This extends a similar
%result in \cite{Emerson&Namjoshi98} for monotone affine function $f(x)=Ax+b$ such that the set
%$\{A^p ; p \geq 0 \}$ is finite (we also say in \cite{FinkelL02} that the monoid
%of the matrix $A$
%is finite; in this case, the set $\cup_{p \geq 0}f^p(x)$ is Presburger-definable hence the lub is also Presburger-definable). Here we show that this last
%hypothesis is not necessary.
We have recalled that the transitions function of Petri nets
($f(x)=x+\vec b$, $\vec b \in \mathbb{Z}^n$ and $\dom(f)$
upward-closed) can be lub-accelerated effectively. This result was
generalized to broadcast protocols (equivalent to transfer Petri nets)
by Emerson and Namjoshi \cite{Emerson&Namjoshi98} and to another class
of monotonic affine functions $f(\vec x)=A \vec x + \vec b$ such that
$A \in \mathbb{N}^{n \times n}$, $b \in \mathbb{N}^n$ (note that $b$
is not in $\mathbb{Z}^n$) and $\dom(f)$ is upward closed
\cite{FinkelMP04}.

\cite{CFS-atpn2011} recently extended this result to
all monotonic affine functions: for every $f(\vec x)=A\vec
x+\vec b$ with $A \in \mathbb{N}^{n \times n}$, $\vec b \in
\mathbb{Z}^n$ and $\dom(f)$ upward-closed, the function $\accel
f$ is recursive. 

%Antonik recently extended this result to
%Presburger monotone affine functions: for every $f(\vec x)=A\vec
%x+\vec b$ with $A \in \mathbb{N}^{n \times n}$, $\vec b \in
%\mathbb{Z}^n$ and $\dom(f)$ Presburger-definable, the function $\accel
%f$ is recursive \cite{AA09}. 

We deduce the following strong
relationship between well-structured ACS and complete well-structured
ACS.

%\begin{texttheorem}{cwscs}
%An affine counter system is a complete WSTS iff it is a WSTS
%with the strong monotony iff all definition domains are upward-closed.
%\end{texttheorem}

\begin{thm}
  \label{cwscs}
  The completion of an $ACS$ $S$ is an $\infty$-effective complete
  WSTS iff $S$ is a strongly monotonic WSTS.
\end{thm}
\proof
  Strong monotonicity reduces to partial monotonicity of each map $f$,
  as discussed above.  Well-structured $ACS$ are clearly effective,
  since $Post(\vec s) = \{\vec t \mid \exists f \in F \cdot f(\vec t)
  = \vec s \}$ is Presburger-definable.
%!!! inclure la suite?
% : there is
%   an algorithm to compute a finite basis $B$ of $\upc Pre^*(\upc \vec
%   s)$.  A basis $B$ of the set $\upc Pre^*(\upc \vec s)$ is equal to
%   $B = \Min A$ where $A=\{\vec t \mid Post(\vec t) \geq \vec s
%   \}=\{\vec t \mid \exists f \in F \cdot f(\vec t) \geq \vec s \}$;
%   $A$ is
%   Presburger-definable, % because all $f$ are Presburger-definable;
%   hence $B$ is computable.
  Note also that monotonic affine function are continuous, and
  $\mathbb{N}_{\omega}^n$ is a continuous dcwo.  Finally, for every
  Presburger monotonic affine function $f$, the function $\accel f$ is
  recursive, so the considered $ACS$ is $\infty$-effective.  \qed

%Hence we may :

\begin{cor}
  One can decide whether the completion of an $ACS$ is an
  $\infty$-effective complete WSTS.
\end{cor}

So the completions of reset/transfer Petri nets \cite{DufourdFS98},
broadcast protocols \cite{Esparza&Finkel&Mayr99}, self-modifying Petri
nets \cite{Valk:smnets} and affine well-structured nets
\cite{FinkelMP04} are $\infty$-effective complete WSTS.

% We will study the completions of B-VASS \cite{VGL-dmtcs05}, Transfer
% Data nets \cite{LazicNORW08}, Reconfigurable nets, Timed Petri nets \cite{AbdullaDMN04}
% Post-Self-modified Petri nets \cite{Valk:smnets} and strongly
% monotone affine well-structured nets \cite{FinkelMP04}, in particular for deciding whether they are clover-flattable.

%%%%%%%%%

%%%%%%%     verifier que la composee de 2 fonctions lineaires monotones est
%%%%%%%       encore pareil
%%%%%%%%%%%
%\section{Finite Representations of Well-Ordered Sets}

%%%%%%%%%%%%%%%%%%%%%%%%%%%%%%%%%%%%%%%%%%%%%%%%%%%%%%%%%%
%%\subsection{Comparison with Adequate Domain of Limits}
%%%%%%%%%%%%%%%%%%%%%%%%%%%%%%%%%%%%%%%%%%%%%%%%%%%%%%%%%%

%Let $S$ be a set of states completed with its ADL $L$. The set $S
%\cup L$ may be considered as a dcpo; moreover, for every $A=\dc
%A \subseteq S \cup L$, $A=\dc F$ with $F \subseteq  S \cup L$ a
%finite set then $\Lub(A)=\dc F$; hence in
%particular, the clover set has still a finite
%basis (this is the main result of theorem $1$). For computing the clover set, we don't need the
%existence of an Adequate Domain of Limits: in fact, not
%\emph{all} downward-closed sets must have a finite basis: it is sufficient
%that every complete downward-closed set, has a finite
%basis. We make a more
%detailed comparison between our framework and the Adequate Domain of
%Limits \cite{GRvB:eec} (See the appendix).

%%%%%%%%%%%%%%%%%%%%%%%%%%%%%%%%%%%%%%%%%%%%%%%%%%%%%%%%%
\section{Conclusion and Perspectives}
%%%%%%%%%%%%%%%%%%%%%%%%%%%%%%%%%%%%%%%%%%%%%%%%%%%%%%%%%

We have provided a framework of {\em complete WSTS\/}, and of {\em
  completions\/} of WSTS, on which forward reachability analyses can
be conducted, using natural finite representations for downward-closed
sets.  The central element of this theory is the {\em clover\/}, i.e.,
the set of maximal elements of the closure of the cover.  We have
shown that, for complete WSTS, the clover is finite and describes the
closure of the cover exactly.  When the original WSTS is not complete,
we have shown the general completion
% [jgl] ``completion'' en double vire' (referee 2)
of WSTS defined in \cite{FGL-stacs2009} is still a WSTS, iff the
original WSTS is an {\em $\omega^2$-WSTS\/}.  This delineates a new,
robust class of WSTS: all known WSTS are $\omega^2$-WSTS.  The
property of being an $\omega^2$-WSTS is also important to ensure
progress in Karp-Miller-like procedures.

We have also defined a simple procedure, {\bf Clover}$_{\mathfrak S}$
for computing the clover for $\infty$-effective complete WSTS
$\mathfrak S$.  This captures the essence of generalized forms of the
Karp-Miller procedure, while terminating in more cases.  We have shown
that that {\bf Clover}$_{\mathfrak S}$ terminates iff the WSTS is {\em
  clover-flattable\/}, i.e., that it is some form of projection of a
flat system, with the same clover.  We have also shown that several
variants of the notion of clover-flattability were in fact equivalent.
We believe that this characterization is an important, and non-trivial
result.

% [jgl] Le referee 1 dit ``In one way or another, it would be nice to recover in your framework the decidability of the coverability problem with a forward approach as it is done in [21]''; je suis d'accord, mais ca ne logera pas (dans la version journal?)

%%%%%%%%%%%%%%%%%%
In the future, we shall explore efficient strategies for choosing
sequences $g \in F^*$ to lub-accelerate in the {\bf
  Clover}$_{\mathfrak S}$ procedure.  We will also analyze whether
{\bf Clover}$_{\mathfrak S}$ terminates in models such as BVASS
\cite{VGL-dmtcs05},
% transfer data nets \cite{LazicNORW08},
reconfigurable nets, timed Petri nets \cite{AbdullaDMN04},
post-self-modifying Petri nets \cite{Valk:smnets} and strongly
monotonic affine well-structured nets \cite{FinkelMP04}), i.e., whether
they are cover-flattable.

One potential use of the clover is in deciding coverability. But the
{\bf Clover}$_{\mathfrak S}$ procedure may fail to terminate.  This is
in contrast to the Expand, Enlarge and Check forward algorithm of
\cite{RB07}, which always terminates, hence decides coverability. One
may want to combine the best of both worlds, and the lub-accelerations
of {\bf Clover}$_{\mathfrak S}$ can profitably be used to improve the
efficiency of the Expand, Enlarge and Check algorithm. This remains to
be explored.

Finally, recall that computing the finite clover is a first step
\cite{Emerson&Namjoshi98} in the direction of solving liveness
properties (and not only safety properties which reduce to
coverability). We plan to clarify the construction of a
\emph{cloverability graph} which would be the basis for liveness model
checking.

\section*{Acknowledgement}
The authors wish to acknowledge fruitful discussions with Sylvain
Schmitz, and to thank the anonymous referees for their comments.

%% in general the use of bibtex is encourage
\bibliographystyle{alpha}
\bibliography{post}

\end{document}